%% file: xuverosion.tex
\documentclass[structabstract]{aa}  
\usepackage{natbib}
\usepackage{graphicx}
\usepackage{lscape}
\usepackage{txfonts}
\begin{document}
   \title{Estimation of the XUV radiation onto close planets and their
     evaporation}

   \author{J. Sanz-Forcada\inst{1}
          \and
          G. Micela\inst{2}
          \and
          I. Ribas\inst{3}
          \and
          A. M. T. Pollock\inst{4}
          \and
          C. Eiroa\inst{5}
          \and
          A. Velasco\inst{1,6}
          \and
          E. Solano\inst{1,6}
          \and
          D. Garc\'{i}a-\'Alvarez\inst{7,8}
          }

   \institute{Departamento de Astrof\'{i}sica,
     Centro de Astrobiolog\'{i}a (CSIC-INTA), ESAC Campus, P.O. Box 78, 
     E-28691 Villanueva de la Ca\~nada, Madrid, Spain; \\
     \email{jsanz@cab.inta-csic.es}
     \and
     INAF -- Osservatorio Astronomico di Palermo
     G. S. Vaiana, Piazza del Parlamento, 1; Palermo, I-90134, Italy%\\
     \and
     Institut de Ci\`ences de l'Espai (CSIC-IEEC), Campus UAB, Fac.
     de Ci\`encies, Torre C5-parell-2$^{\underline a}$ planta, 
     E-08193 Bellaterra, Spain%\\
     \and
     XMM-Newton SOC, European Space Agency, ESAC, Apartado 78,
     E-28691 Villanueva de la Ca\~nada, Madrid, Spain%\\ 
     \and
     Dpto. de F\'{i}sica Te\'orica, C-XI, Facultad de Ciencias, 
     Universidad Aut\'onoma de Madrid, Cantoblanco, E-28049 Madrid, Spain%\\
     \and
     Spanish Virtual Observatory, Centro de Astrobiolog\'{i}a
     (CSIC-INTA), ESAC Campus, Madrid, Spain%\\
     \and
     Instituto de Astrof\'{i}sica de Canarias, E-38205 La Laguna, Spain %\\
     \and
     Grantecan CALP, E-38712 Bre\~na Baja, La Palma, Spain%\\
              }

   \date{Received ; accepted }

% \abstract{}{}{}{}{} 
% 5 {} token are mandatory
 
  \abstract
   {The current distribution of planet mass vs.
     incident stellar X-ray flux supports the idea that 
     photoevaporation of the atmosphere may take place in close-in
     planets. Integrated effects have to be accounted for. A proper
     calculation of the mass loss rate due 
     to photoevaporation requires to estimate the total
     irradiation from the whole XUV range.}
   {The purpose of this paper is to extend the analysis of the
       photoevaporation in planetary atmospheres from 
     the accessible X-rays to the mostly unobserved EUV range by
     using the coronal models of stars to calculate the EUV
     contribution to the stellar spectra. The mass evolution of
     planets can be traced assuming 
     that thermal losses dominate the mass loss of their atmospheres.}
   {We determine coronal models for 82 stars with exoplanets that
     have X-ray observations available. Then a synthetic spectrum is
     produced for the whole XUV range ($\sim$1 -- 912 \AA). The
     determination of the EUV stellar flux, calibrated with real EUV
     data, allows us to calculate the
     accumulated effects of the XUV irradiation on the planet
     atmosphere with time, as well as the mass evolution for planets
     with known density.}
   {We calibrate for the first time a relation of the EUV luminosity
     with stellar age valid for late-type stars.   
     In a sample of 109 exoplanets, few planets with masses
     larger than $\sim$1.5 M$_{\rm J}$ receive high XUV flux,
     suggesting that intense photoevaporation takes place in a short
     period of time, as previously found in X-rays. The
     scenario is also consistent with the observed distribution of
     planet masses with density. The accumulated effects of 
     photoevaporation over time indicate that HD~209458b may have lost 
     0.2 M$_{\rm J}$ since an age of 20~Myr.}    
   {Coronal radiation produces rapid photoevaporation of the
     atmospheres of planets close to young late-type stars. More
     complex models are needed to explain fully the
     observations. Spectral energy distributions in the XUV 
     range are made available for stars in the sample through the Virtual
     Observatory, for the use in future planet atmospheric models.}

   \keywords{(stars:) planetary systems -- stars: coronae --
     astrobiology -- x-rays: stars}

   \maketitle
%
%________________________________________________________________

%vvvvvvvvvvvvvvvvvvvvvvvvvvvvvvvvvvvvvvvvvvvvvvvvvvvvv
\section{Introduction}
After 15 years of exoplanetary science, the discipline has reached a point
at which it is possible to study in more detail the physical
properties of planets, their formation and evolution. 
With more than 500 exoplanets known to date it
is possible to explore relations between the planets and
their host stars. In particular, the mass of the planets and the
atmospheric conditions are partly linked to the stellar radiation.
Once the planet is formed, and the original disc is
dissipated, the main agent interacting with the atmosphere should be
the high energy emission from the corona of the star, for late type
stars \citep[Sanz-Forcada et al. 2010b, and references
  therein]{lam03,erk07,pen08b,cec09}. 
Recently Sanz-Forcada et al. (2010b), \citep[hereafter][]{san10} noticed
that erosion in exoplanets may be 
taking place as an effect of coronal radiation vaporizing the
atmosphere of close-in planets: they showed that
distribution of planet masses with the X-ray flux at the planet
implies that the least massive planets currently receive a high X-ray flux.
This is
interpreted as the effect of radiation erosion in the long term.

Photons with $\lambda < 912$~\AA\ can ionize hydrogen atoms, 
assumed to be the main component of the atmosphere of giant
planets. The effects of X-rays ($\lambda\lambda$~5--100) and
Extreme Ultraviolet (EUV, $\lambda\lambda$~100--912) photons take
place at different heights in the atmosphere of the 
planet. While EUV photons mainly ionize the atoms in the upper
atmosphere, X-rays penetrate deeper into the atmosphere. The
free electrons carry a high momentum, producing a cascade of
collisions while the X-rays 
photons are absorbed in the atmosphere \citep{cec09}. These
collisions heat the 
atmosphere yielding its ``inflation'' and
eventually the evaporation of a part of it. The gravity of the planet
acts as protection, trying to keep the atmosphere attached to the
planet. If we assume that the planet atmosphere is mainly composed of
hydrogen, and all the photons in the whole XUV range (X-rays+EUV) are
absorbed and contribute to the heating of the atmosphere, it is
possible to calculate the mass loss of the atmosphere by balancing the
losses with the planet gravity \citep{wat81,lam03}. An additional source of
mass loss takes place through the Roche Lobe for close-in planets
\citep{erk07} \citep[see also][for detailed
  simulations of the escape through Roche Lobe]{lec04,jar05}. 
The resulting formula \citep{san10} is:

\begin{equation}\label{eq:general}
\dot M=\frac{\pi R_{\rm p}^3 F_{\rm XUV}}{{\rm G} K M_{\rm  p}}
\end{equation}
where $K$ ($K \le 1$) accounts for the planet radius Roche Lobe losses
\citep{erk07}, $F_{\rm XUV}$ is the X-ray and EUV flux at the planet orbit, 
and G is the gravitational constant\footnote{A factor of 4, used in \citet{pen08b} and \citet{san10}, has been removed to account for geometrical considerations, as explained in Eq.~21 of \citet{erk07}}. 
\citet{bar04} consider that evaporation actually takes
place at a point somewhere above the planet radius $R_{\rm p}$, at the
``expansion radius'' $R_1$. However the bulk of the XUV radiation should
be absorbed by material enclosed within the planet radius, so we
assume that $R_1 \simeq R_{\rm p}$.
The formula can be simplified using the
mean density of the planet ($\rho$), and assuming that $K \simeq 1$
(valid for most cases, and a lower limit to the mass loss in any case):

\begin{equation}\label{eq:massloss}
\dot M=\frac{3 F_{\rm XUV}}{4\, {\rm G} \rho}\, .
\end{equation}

\input{tabobslog}

Stars in the range late-F to mid-M stellar types are characterized by
coronae with temperatures of $\sim$1~MK, exceeding 10~MK in the most
active cases. The high temperature material in the transition region
($\sim \log T=4-5.8$) and corona ($\sim \log T=5.8-7.4$) emits
copious X-rays and EUV flux. Fast rotators have hotter coronae,
resulting in higher XUV fluxes.
Current astronomical instruments give 
access to the X-rays band only. More energetic flux ($\lambda \la$1~\AA) is
negligible even for active stars, while the radiation in the EUV band
is severely absorbed by the 
neutral and molecular hydrogen in the interstellar medium (ISM).
There are no 
missions currently observing in the EUV band, and the few data from past
telescopes, such as EUVE (Extreme Ultraviolet Explorer), are limited to the
closest stars in the range $\lambda\lambda 100-400$. Only one star
hosting an exoplanet has been observed in this band, $\epsilon$~Eri
\citep{san03b}.

\input{tabrosatlog}

\input{tabfits}

The present-day distribution of planet masses 
should reflect the accumulated effects
of mass loss in the atmospheres of the planet over time
\citep{san10}. Moreover,  if we know the evolution of
the emission in the whole XUV band we 
should be able to trace the planet evolution, given an accurate
knowledge of the density of the planet, according to
Eq.~\ref{eq:massloss}.
Since younger stars have faster
rotation, their coronae emit more XUV flux, which therefore
decreases with time. The evolution of the X-ray
emission with age has been 
studied for the Sun \citep[e.g.][]{mag87,ayr97,rib05} and extended to
G and M stars \citep{pen08a,pen08b}.
A relation using late F to early
M stars has been calibrated by Garc\'es et al. \citetext{in preparation} 
in the X-ray band,
allowing us to calculate the age of the stars from its X-ray
emission \citep{san10}, and to trace the time evolution of this
emission. In the EUV band the time evolution of the emission has been only
studied for the Sun \citep{rib05}.

In this work we extend the analysis of \citet{san10} to the EUV band,
to account for all  
the stellar radiation capable to 
ionize hydrogen in a planet atmosphere. \citet{lec07} extrapolated the
EUV flux from the X-ray flux, using the Sun as pattern for all kind
of stars, without checking whether the relation is
valid at all levels of activity, and therefore ages. 
Since there are essentially no measurements in 
the range $\sim$400--912 \AA\ for stars other than the Sun, we use
coronal models to synthesize the Spectral Energy Distribution in the
whole EUV range, and test the results in X-rays and the lower
wavelengths of the band (100--400~\AA) for a few cases with EUV spectra
available. We have set up a database
(http://sdc.cab.inta-csic.es/xexoplanets), that is freely 
available (within the Spanish Virtual Observatory), ``X-exoplanets''. 
The database
includes synthetic SEDs in the range 1--1200~\AA\ for all the stars
listed in Table~\ref{tabobslog}. Objects will be incorporated in the
future as they are observed in X-rays.

The paper is structured as follows: Section 2 describes the
observations, Sect. 3 extends the X-ray analysis to the EUV band;
Sect. 4 show the results found for the sample, that are discussed in
Sect. 5, to list the conclusions of the work in Sect. 6.

%vvvvvvvvvvvvvvvvvvvvvvvvvvvvvvvvvvvvvvvvvvvvvvvvvvvvv
\section{Observations}
Data acquired with the X-ray telescopes XMM-Newton, Chandra and ROSAT
have been used in this work (Tables~\ref{tabobslog},~\ref{tabrosatlog}). 
XMM-Newton and Chandra data were
fetched from their public archives, including data awarded to us as
P.I. or co-I. (XMM prop. ID \#020653, \#020000, \#055102).
Data were reduced following standard procedures,
removing time intervals affected by high background, likely
produced by space weather events. The expected position of the targets
were calculated using the coordinates and proper motions provided by
SIMBAD. XMM-Newton/EPIC and Chandra/ACIS have spatial resolution of
6\arcsec\ and 2\arcsec\ respectively.
The cleaned observations (Table~\ref{tabobslog}) were used to extract the 
low-resolution spectra provided by XMM-Newton/EPIC
and Chandra/ACIS (E/$\Delta$E=20--50). 
The ISIS package \citep{isis} and the Astrophysics 
Plasma Emission Database
\citep[APED,][]{aped} were used to fit the spectra with
coronal models of 1 to 3 temperature components (Table~\ref{tabfits}),
depending on the 
quality of the spectra, and variable stellar coronal abundances, using 
the generally small value of the ISM absorption. The background
spectrum was fitted simultaneously to the source to account for its
contribution to the total spectrum. Spectra and light curves for each
target are
available on-line (http://sdc.cab.inta-csic.es/xexoplanets) in
the ``X-exoplanets'' database \citep{san09}, described in detail in
Appendix~\ref{app:xexo}. Spectra in the XUV range can be used for
planet atmospheric models.

%----------------------------------  Fig. 1
   \begin{figure}[t]
   \centering
   \includegraphics[width=0.5\textwidth,clip]{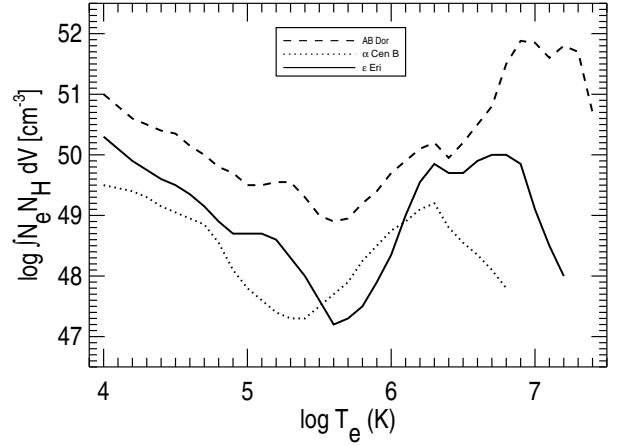}
   \caption{Emission Measure Distribution (EMD) of three K2V stars
     with different levels of activity. The shape of solar EMD
     \citep{orl01} is
     similar to that of $\alpha$~Cen B.}\label{EMDs}
    \end{figure}
%----------------------------------------------
%

The spectra with lowest statistics have a deficient fit. This
yields low abundance and
very high emission measure, despite of the low temperature
observed (emission measure in stellar coronae increase with the
  their coronal temperature).
The use of solar abundance provides a similar fit, but with more
realistic values of the emission measure. In these cases we
fixed temperature and abundance to the solar values. We use $\log
T$\,[K]=6.3 and the solar photospheric abundances of
\citet{asp05},  
corresponding to [Fe/H]=--0.2 in the scale of
\citet{anders} used in Table~\ref{tabfits}. The actual
model used in the fit has little influence in the calculation of the
X-ray (0.12--2.48~keV or $\sim$5--100\AA) flux displayed in
Table~\ref{tabfits}, but it is 
important for the extension to the EUV range, as explained in
Sect.~\ref{sec:euv}.
Table~\ref{tabfits} lists also the errors for objects with
net count rates with S/N$>$3, and consider the rest of detections as upper
limits. We mark GJ~86 as an upper limit to account for the contribution
of an unresolved companion.

ROSAT/PSPC observations were added to the sample. We consider
only detections with S/N$>$3, given the lower spatial resolution of
this instrument (25\arcsec), 
marking as upper limits objects with suspected X-ray bright
companions, as indicated in Table~\ref{tabrosatlog} (as a reference, a
dM3 star may have up to $\log L_{\rm X} \sim 27.5$). To calculate the
X-ray flux of the targets we consider the count rate ($CR$) reported
in HEASARC (http://heasarc.gsfc.nasa.gov/), 
corresponding to the spectral range 0.12--2.48~keV, and then
transform it into flux using 
$f_{\rm X}=CR \times 6.19\times 10^{-12}$\,erg\,cm$^{-2}$\,cts$^{-1}$, 
as proposed by
\citet{sch95} and \citet{hue98}, using a hardness ratio of --0.4
(corresponding to a middle 
activity level star, $\epsilon$~Eri). 

The stellar distance, not available for three ROSAT stars,
was calculated using the spectroscopic parallax method, i.e., comparing the
visual magnitude V with the absolute magnitude
M$_{\rm V}$ that corresponds to the spectral type of the star
\citep{cox00}. We calculated the bolometric luminosity
($L_{\rm bol}$) of each star using the bolometric corrections by
\citet{flo96} and B-V colors based on spectral types in \citet{cox00} if no
direct measurements are available in SIMBAD. The calculation of the age of the
stars depends partly on $L_{\rm bol}$ \citep{san10}. We notice
discrepancies with the $L_{\rm bol}$ calculated by \citet{pop10} for
some objects. The latter values are unrealistically high for GJ 832
(M1, $\log L_{\rm 
  bol}$ [erg s$^{-1}$]=35.86), GJ 176/HD 285968 (M2V, 33.61), GJ 436 (M2,
33.52), HR 810 (G0V, 36.19), GJ 3021 (G6V, 36.02), HD 87833 (K0V,
34.37). These values 
correspond to spectral types of more massive stars. Finally
some X-ray fluxes differ from other surveys \citep{kas08,pop10} mainly
because they calculated the X-ray flux as a count rate conversion for
most objects, or because of misidentifications of the stellar position
in the field, as described in \citet{san10}.

%----------------------------------  Fig. 2
   \begin{figure}[t]
   \centering
   \includegraphics[width=0.49\textwidth,clip]{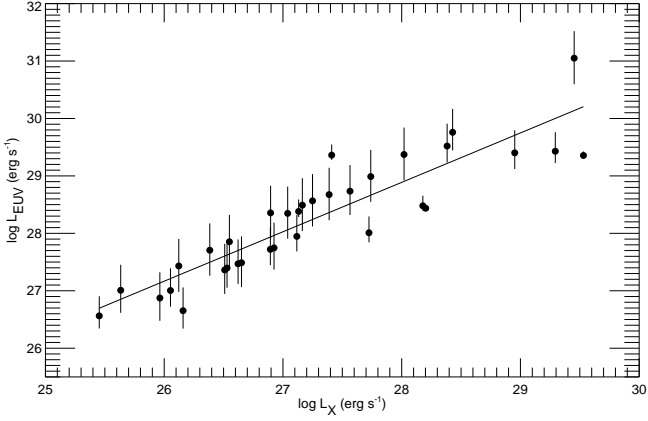}
   \caption{X-ray vs EUV luminosity in selected stars of the sample
     (see text). Uncertainties come from calculation of the
     transition region EMD.}\label{relxeuv}
    \end{figure}
%----------------------------------------------
%

%vvvvvvvvvvvvvvvvvvvvvvvvvvvvvvvvvvvvvvvvvvvvvvvvvvvvv
\section{Extension to the EUV}\label{sec:euv}
In \citet{san10} we calculated the X-ray fluxes that were received at
the planet orbit ($F_{\rm X}$). We need to consider all the flux
emitted by the star in the X-ray and EUV range in order to calculate the mass
loss rate of the planet according to Eq.~\ref{eq:massloss}.  
\citet{lec07} extrapolated the EUV flux from the X-ray-to-EUV
ratio observed in the Sun, assuming the same ratio
for all spectral types and activity levels (thus age). The few
observations of late type stars in the EUV contradict this view
\citep[e.g.][]{san03b}. 
A better way to determine the radiation in the EUV range is to calculate a
synthetic spectrum of each star using a coronal model.
Such a model must describe accurately how the mass is
distributed with 
temperature in the corona and transition region, the so-called
Emission Measure Distribution (EMD). The EMD is complemented with the
abundances of the different elements in the corona. 
If combined with an appropriate atomic
model, it will be possible to predict the spectral energy distribution
generated at coronal temperatures. Atomic
models in this range have been tested only with solar data, and
at the shorter wavelengths ($\lambda\lambda\sim 100 - 400$~\AA) with
more active stars. We expect that a few lines formed at high
temperature have inaccurate calculations or are missing from the
models. But the bulk of the emission is already included in the
models and the flux in a given range should be quite accurate. We use
in our case the atomic model APED \citep{aped}.  

%----------------------------------  Fig. 3
   \begin{figure}[t]
   \centering
   \includegraphics[width=0.49\textwidth]{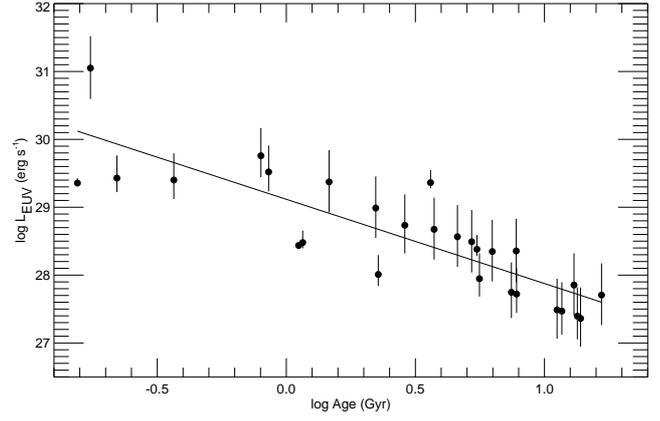}
   \caption{Distribution of EUV luminosities against age (in Gyr)
     determined using the X-ray luminosities. The line indicates the
     best linear fit to the sample. }\label{fig:evoleuv}
    \end{figure}
%----------------------------------------------
%

\input{tabbands}

\input{tabfluxrosat}
\input{tabfluxes}

The atomic model needs to be folded with a coronal
model. Fig.~\ref{EMDs} shows the EMDs of three K2V stars with good
high-resolution spectra, that allow us to construct an EMD with a
resolution of 0.1 dex in temperature, following the method described in
\citet[and references therein]{san03a}. In
general we do not have an accurate determination of the EMD for the
stars in the sample, most of them limited to fits with less than 3
temperature components. Moreover, the observations in X-rays give
us access only to the EM at T$\ga$1 MK,
since lower temperature lines are rare in X-rays. However we cannot
ignore the lower temperature part of the EMD, which corresponds to the
transition region, because many lines in the
EUV band are formed at T$\la$1 MK. It is remarkable the case of the cooler
\ion{He}{ii}~304~\AA\ line, one of the
strongest in the XUV spectrum of most cool stars.

\subsection{Transition region EMD}
It is possible to calculate the EMD at T$\la$1~MK by using lines
in the UV \citep[e.g.][]{dup93}, but most of our objects have no UV
spectra available. 
The large sample of EMDs reported by
\citet{san03b} showed that the ``cool'' side of the EMD is
 approximately proportional to the EMD at $\log T$\,(K)$\sim6-6.3$. We checked
that a reasonable proportionality exists among the values at these two
temperature ranges (Appendix~\ref{loweremd}). 
We have calibrated the relation using stars with
known EMD at all temperatures, and extrapolated the lower temperature
EMD, for the stars unobserved in UV, using this relation, as described
in Appendix~\ref{loweremd}.
In particular we chose the EMD of the sample of stars in
\citet{san03b} that have well calculated EMD: we separated the sample
in three groups depending on the level of activity (interpreted from
the amount of EMD found at the highest temperatures).  
We define then
three parameters: a parameter to account for the transformation
between an EMD with 0.1~dex of temperature resolution and a 3T fit, the
difference between the minimum of the EMD and the local maximum found at
$\log T$\,(K)$\sim6.2-6.4$, and the slope of the EMD at temperatures below the
EMD minimum. We have then tested this extrapolation with one star of
each of the three classes: AB Dor \citep{san03a}, $\epsilon$~Eri
\citep{san04}, and $\alpha$~Cen~B (Sect.~\ref{sec:highres}), all of them of
type K2V (Fig.~\ref{EMDs}). 
The XUV flux calculated using a 3-T model ($\log T$\,(K)$\ge 5.8$) and a
synthetic EMD ($\log T$\,(K)$<5.8$), agrees in all cases within 0.37
dex with that 
calculated using a complete EMD. Therefore we are confident that the 
method can be safely applied to the stars in our sample. As a
comparison, the solar activity cycle spans 1.7 dex in X-ray luminosity
\citep{orl01}. We use also these stars to check how much
flux we would 
be missing if we consider the EMD only above 1\,MK, for the
different levels of activity: we would miss a 3\% of the flux (AB Dor,
very active), 19\% (intermediate activity $\epsilon$~Eri), and 40\%
($\alpha$~Cen~B, low activity). We conclude that the
extension of the EMD to the cooler temperatures is necessary, especially
for the less active stars. The reason is because in  $\alpha$~Cen~B
91\% of the XUV flux is generated in the EUV band, while for AB Dor only
23\% is in the EUV. Table~\ref{tab:bands} include the predicted flux in
different bands of the EUV range, with a formal precision of 0.01
  dex well below the expected cyclic variability of the stars.

The application of this calculation to the ROSAT data is not
possible since we do not have a coronal model for these
objects. Instead we calculated a direct relation between X-ray
and EUV luminosities for all the XMM and Chandra objects in our sample 
that have an X-ray flux available (i.e., it is not an upper
limit), with the EUV flux calculated as explained. 
A fit to these data (Fig.~\ref{relxeuv}) yields:

\begin{equation}
\log L_{\rm EUV}= (4.80 \pm 1.99) + (0.860 \pm 0.073) \log L_{\rm X}
\end{equation}
where $L_{\rm X}$ and $L_{\rm EUV}$ are the X-ray ($\lambda\lambda$
5-100~\AA) and EUV 
($\lambda\lambda$ 100-920~\AA) luminosities, in erg\,s$^{-1}$. We have
then applied this conversion to the ROSAT data. The calculated EUV
luminosities are listed in Table~\ref{tabfluxrosat}.

\subsection{EUV luminosity evolution with age}
We have calibrated a relation of age-$L_{\rm EUV}$ to calculate better
the effects of the radiation in the whole XUV range. We have
selected for the calibration all the stars in our sample, excluding
those with only an upper limit of the X-ray flux, and stars
that show a formal result with age larger than 20 Gyr (those with
spectral type A and some M stars). The age determination is made with
the X-ray luminosity as explained in next Section. The fit  
(Fig.~\ref{fig:evoleuv}) follows a power-law relation:  

{\setlength\arraycolsep{2pt}
\begin{eqnarray}\label{eq:laweuvage}
\log \,L_{\rm EUV} & = & (29.12 \pm 0.11)\, -(1.24\pm0.15) \log \tau 
\end{eqnarray}}
where $\tau$ is the age in Gyr. This relation has been
used to calculate the accumulated effects of the coronal radiation in
the planet atmosphere as listed in
Tables~\ref{tabfluxrosat},\,\ref{tabfluxes}.
Although \citet{rib05} studied the time evolution of the EUV
luminosity, this is the first time
that a general relation age-$L_{\rm EUV}$ has been calculated.
\citet{rib05} studied the spectral energy distributions of six solar
analogs covering a wide range (0.1--6.7 Gyr) of stellar ages, in the
X-rays and UV regimes. They used real spectra in the ranges 1--360~\AA\
and 920--1950~\AA\ as available. The distribution of stellar fluxes
in the different bands with the stellar age was used to calibrate the
time evolution of the high energy irradiance in these bands. The
results were used to interpolate the evolution
in the 360--920~\AA\ band, resulting in a less steeper
dependence (exponent $\sim -1.0$ in this range, $-1.20$ in 100--360~\AA),
than in our case ($-1.24$ in 100--920~\AA).

%vvvvvvvvvvvvvvvvvvvvvvvvvvvvvvvvvvvvvvvvvvvvvvvvvvvvv
\section{Results}
Processes of mass loss in the atmosphere of a planet are not well
understood. Once the circumstellar disc is dissipated it is expected that
thermal losses are dominant. The mass loss rate due to thermal losses
in a planet with an atmosphere mostly composed by hydrogen is defined
by Eq.~\ref{eq:massloss}.
To account for all the energy
budget in the XUV band we need to know the radiation that is absorbed
by neutral hydrogen, i.e., photons with $\lambda < 912$~\AA. In
\citet{san10} we considered only the X-rays photons as a proxy of the
whole XUV radiation. It is thus necessary to test whether the
conclusions achieved hold also for the whole XUV range. With all the
energy budget we can make a first estimate of the total mass lost
due to thermal losses for the planets in the sample with known
density (see below). Figures~\ref{masses} and \ref{masses2} show the
distribution of planet masses ($M_{\rm P} \sin i$) with the X-ray
flux at the planet orbit (directly measured) and the XUV flux that we
have calculated after modeling of corona and transition region. 
A dashed line ($\log F_{\rm X}=3-0.5$\,$M_{\rm p} \sin i$) is plotted
  in Figure~\ref{masses} to indicate the ``erosion line'', proposed in
  \citet{san10} to separate what could be a regime of fast erosion from a slower
  erosion phase, based on the observed distribution.
Fig.~\ref{masses2} is a direct indication of the
current mass loss rate in exoplanets, assuming that thermal losses are
dominant and the density is the same for all planets (according
  to Eq.~\ref{eq:massloss}). 
Regrettably we know the
planet density of only four planets in our sample
(Table~\ref{tab:roche}). Rocky 
planets ($\rho \sim 5$\,g\,cm$^{-3}$) are not supposed to be affected in the same
manner (a gaseous planet like Jupiter has $\rho$=1.24).
We marked in the plot the XUV flux that would have arrived at the
Earth's orbit at an age of $\sim$100~Myr and $\sim$1~Gyr. To calculate
this flux we used two young solar analogs, $\kappa$~Cet and EK~Dra, as
explained in \citet{san10}. The emission in the EUV range was
calculated by Sanz-Forcada et al. (in prep.)
extrapolating from a coronal model determined with high
resolution spectra.  

%----------------------------------  Fig. 4
   \begin{figure}[t]
   \centering
   \includegraphics[width=0.49\textwidth,clip]{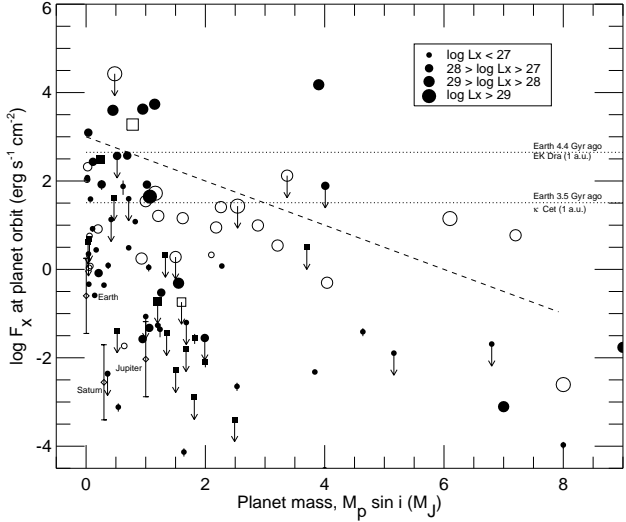}
   \caption{Distribution of planetary masses ($M_{\rm p} \sin i$) with X-ray
     flux at the planet orbit. Filled symbols (squares for subgiants,
     circles for dwarfs) are
     XMM-Newton and Chandra data. Arrows indicate upper
     limits. Open symbols are ROSAT data without error
     bars. Diamonds represent Jupiter, Saturn, and the Earth. 
     The dashed line marks the ``erosion line'' that might separate
     the phases of strong and weak evaporation \citep{san10}. 
     Dotted lines indicate the X-ray flux of the younger Sun
     at 1 a.u.}\label{masses}
    \end{figure}
%----------------------------------------------
%

A better way to measure the long-term effects of the radiation is
to calculate the accumulated XUV flux at the orbit of the planet. To
do that we need to know how coronal radiation evolves with
time. Several laws have been reported in the past, mostly devoted to
explain the coronal history of the Sun as a law of the kind 
$L_{\rm X}\sim t^\alpha$, calibrated with G dwarfs. Some examples are
\citet[][$\alpha =-1.5$]{mag87}, \citet[][$\alpha =-1.74$]{ayr97},
\citet[][$\alpha =-1.5$]{gud97}, \citet[][$\alpha =-1.27$ or $-1.92$,
  for 1--20~\AA\ and 20--100~\AA\ respectively]{rib05},
\citet[][$\alpha=-1.69$]{pen08b}, and \citet[][$\alpha=-1.34$,
  calibrated with M dwarfs]{pen08a}. We use the law by
Garc\'es et al. (in prep.), calibrated with late F to early M dwarfs:
{\setlength\arraycolsep{2pt}
\begin{eqnarray}\label{eq:lawxage}
L_{\rm X} & = & 6.3\times10^{-4}\, L_{\rm bol}  \quad \qquad  (\tau < \tau_i)
\nonumber\\
L_{\rm X} & = & 1.89\times10^{28}\, \tau^{-1.55} \qquad  (\tau > \tau_i)
\end{eqnarray}}
with $\tau_i = 2.03\times 10^{20} L_{\rm bol}^{-0.65}$.  
$L_{\rm X}$ and $L_{\rm bol}$ are in erg\,s$^{-1}$, and $\tau$ is the
age in Gyr. The $\tau_i$ parameter marks the
typical change from saturation regime to an inverse proportionality
between $L_{\rm X}/L_{\rm bol}$ and rotation period
\citep[e.g.][]{piz03}. The relation shows a similar behavior to former
calibrations, but it can be applied to a wider range of stellar
masses. We use Eq.~\ref{eq:lawxage} to calculate the stellar age
(Table~\ref{tabfluxes}), with the caveat that there 
is an uncertainty of about an order of magnitude in the $L_{\rm X}$ levels
of stars of the same spectral type and age \citep{pen08b,pen08a}, and
that the solar cycle in X-rays spans as much as 1.7 dex in 
$L_{\rm X}$ \citep{orl01}. We consider a formal upper limit of the age
at 15~Gyr \citep[current estimate of the age of the Universe is 13.7
Gyr,][]{ben03}. This method is not optimal for the age determination
of a star, but it is more appropriate if we want to know the ``X-ray
age'' in the evolution of the coronal emission of a given star.

%----------------------------------  Fig. 5
   \begin{figure}[t]
   \centering
   \includegraphics[width=0.49\textwidth,clip]{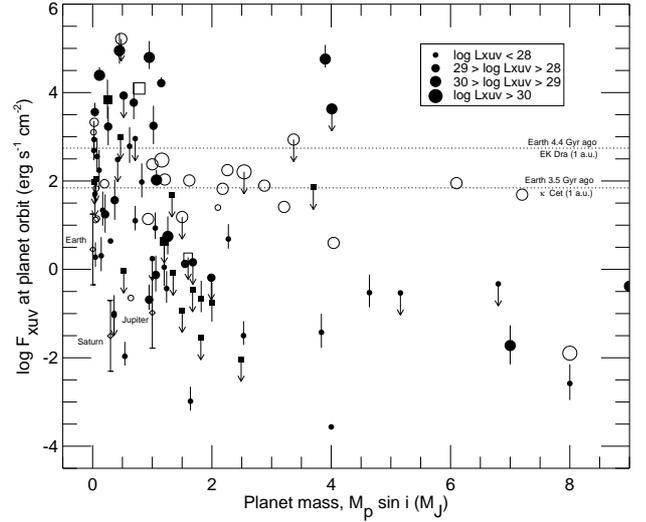}
   \caption{Distribution of planetary masses ($M_{\rm p} \sin i$) with
     XUV flux at the planet orbit. Symbols as in Fig.~\ref{masses}
     Error bars indicate the limits in EUV flux calculated
     with the models.
     Dotted lines indicate the X-ray flux of the younger Sun
     at 1 a.u.
     Note that mass loss rates increase with the XUV flux.
   }\label{masses2}
    \end{figure}
%----------------------------------------------
%

\begin{figure}[t]
   \centering
   \includegraphics[width=0.49\textwidth,clip]{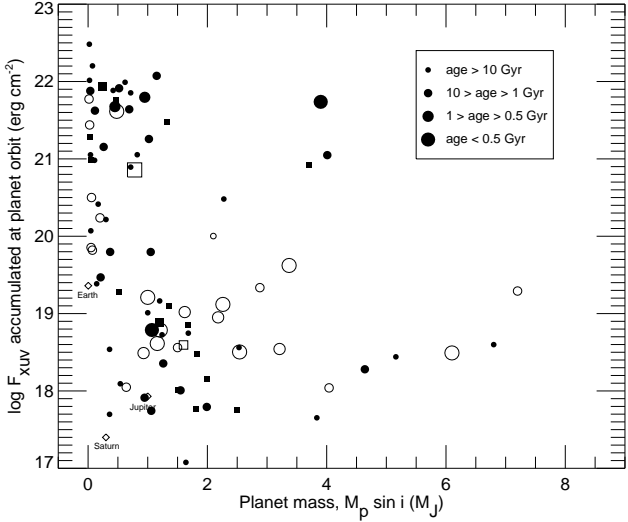}
   \caption{Distribution of planetary masses ($M_{\rm p} \sin i$) with the XUV
     flux accumulated at the planet orbit since an age of 20 Myrs to the
     present day (see text). Symbols as in Fig.~\ref{masses}.
     Note that increasing XUV flux accumulated indicates a larger mass
     lost to date.}\label{agemasses}
    \end{figure}

The accumulated X-ray and
XUV flux at the planet orbit (Table~\ref{tabfluxes},
Fig.~\ref{agemasses}) is calculated using
Eqs.~\ref{eq:laweuvage} and \ref{eq:lawxage}, between 20~Myr and the
present. Most protoplanetary discs would have dissipated after 20~Myr. 
We do not calculate the age for giants and A-type stars, and mark the
subgiants with different symbols since it is not known whether
they follow the same relation, but sometimes the determination of
the star as dwarf or subgiant is not precise. Only planets with $M
\sin i < 9$\,M$_{\rm J}$ are considered for further calculations.
According to Eq.~\ref{eq:massloss},
Fig.~\ref{agemasses} is a direct indication of the total mass lost
to date, assuming same density for all planets. A planet like
$\tau$~Boo~b, with 10$^{21.7}$~erg\,cm$^{-2}$
accumulated in the XUV band during 350 Myr, would have lost
$\sim$0.04~M$_{\rm J}$ if the density is 1\,g\,cm$^{-3}$, but as much as
$\sim$0.7~M$_{\rm J}$ if the density is 0.1\,g\,cm$^{-3}$, among the
lowest observed in exoplanets.

%----------------------------------  Fig. 6
%----------------------------------------------

\subsection{Accumulated effects in planets with known density}
\begin{figure}[t]
  \centering
  \includegraphics[angle=90,width=0.49\textwidth,clip]{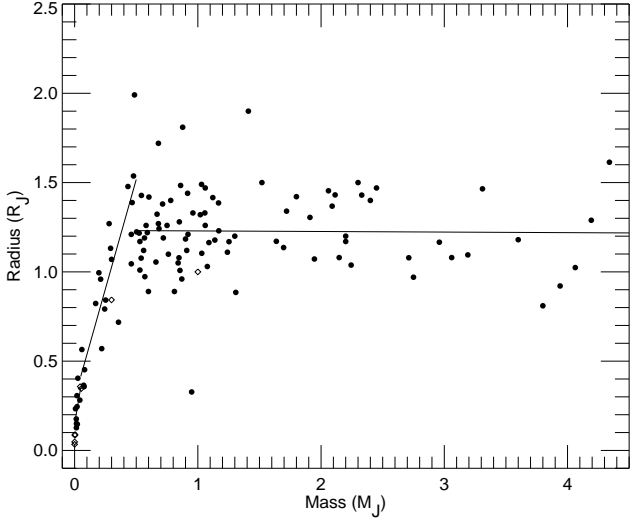}
  \caption{Mass-radius relation in the sample of exoplanets (8 Mar 2011)
    of known radius (filled circles). Open diamonds 
    represent the Solar System planets. Solid lines indicate the
    least-square fit to the data (see text). 
  }\label{fig:massradius}  
\end{figure}

It is expected that the accumulated effects of the XUV radiation
result in a population of close-in planets with less massive planets,
unless they have higher densities, or they simply have little or no
atmosphere. The current distribution of planets with known density is
dominated by close-in planets, 
and therefore it has an appropriate
bias to test these effects, even if we do not know the radiation of
their parent stars. It is remarkable to observe in such distribution
that there are no massive planets with low density at short distances
of the star \citep{san10}. We do not know now if this is
also applicable for planets at further distances, that would indicate
that this is an effect of planet formation. Theoretical models
\citep[e.g.][]{gui05,for07} indicate that irradiated planets might
have a nearly constant radius for planets with $M_{\rm P}\ga
1$~M$_{\rm J}$, explaining the increasing density with mass
in Fig.~3 of \citet{san10}. The radius-mass relation of the same sample
supports this idea (Fig.~\ref{fig:massradius}), although actual values of
planet radius are slightly higher on average than Fig.~10 of
\citet{gui05}. This empirical relation can be defined as follows:
{\setlength\arraycolsep{2pt}
\begin{eqnarray}\label{eq:massradius}
R_{\rm P} & = & (0.15\pm0.07) + (4.1\pm3.0) M_{\rm P} \quad \qquad (M_{\rm
  P} < 0.05) \\
R_{\rm P} & = & (0.29\pm0.11) + (2.45\pm0.36) M_{\rm P} \qquad (0.05 <
M_{\rm P} < 0.5) \nonumber\\
R_{\rm P} & = & (1.23\pm0.05) + (0.00\pm0.02) M_{\rm P} \qquad (0.5 <
M_{\rm P} < 4.5) \nonumber\\ \nonumber
\end{eqnarray}}

\begin{figure}[t]
   \centering
   \includegraphics[width=0.49\textwidth,clip]{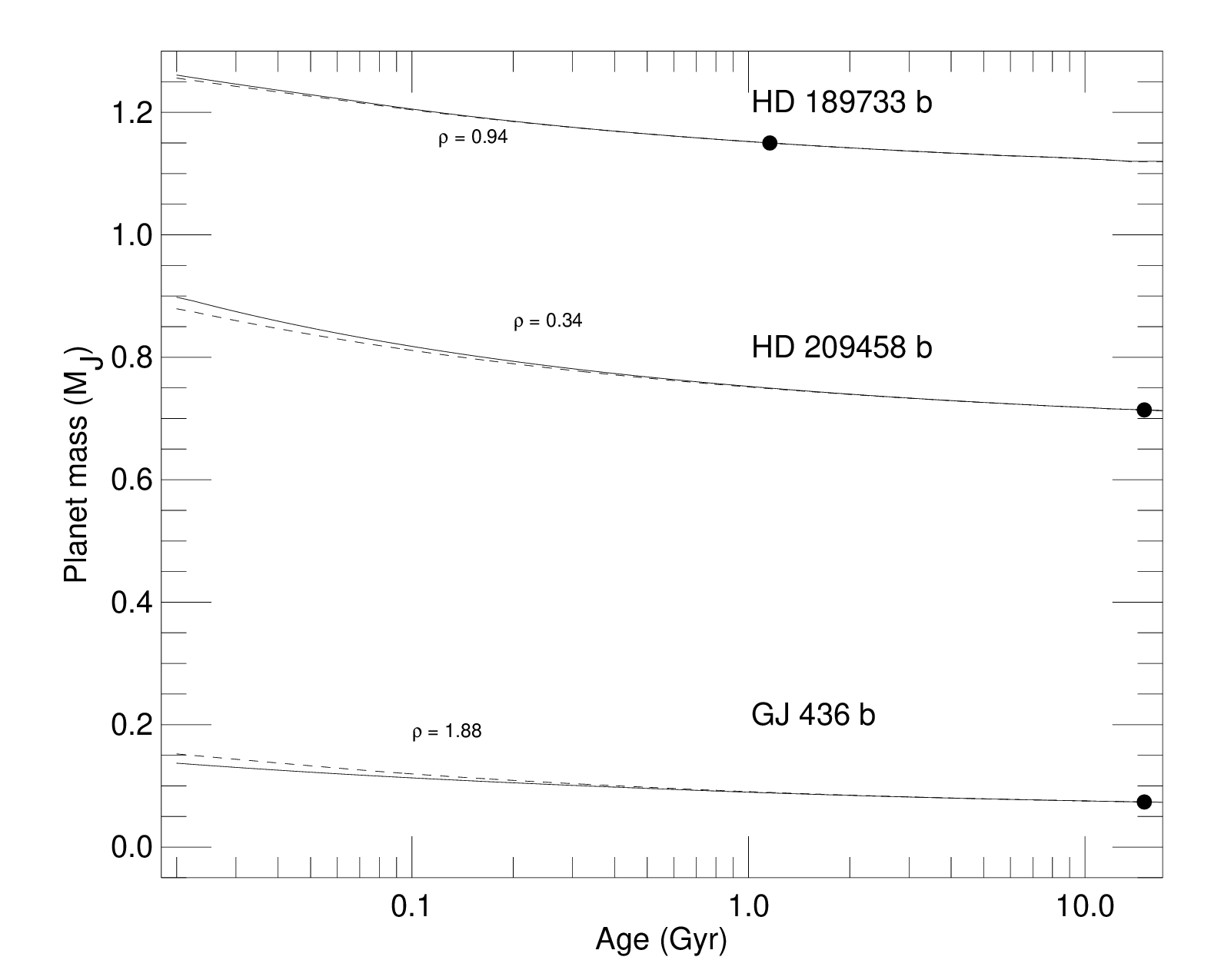}
   \caption{Time evolution of planetary mass, assuming thermal and
     Roche lobe losses. Planet density is indicated. Solid lines
       indicate evolution under constant density conditions, while
       dashed lines consider the radius-mass relation as explained in
       the text (constant radius if $M_{\rm P}>0.5$~M$_{\rm
         J}$). Current age is marked with solid circles.
   }\label{masslosshistory}  
    \end{figure}

With the estimate of the total XUV flux it is possible now to check the
mass loss history 
of the few planets in our sample with known density (we exclude the
A5V star HR~8799). While 2M1207\,b
suffers basically no erosion (Table~\ref{tabfluxes}), the other three
planets (GJ 436 b, HD 189733 b, HD 209458 b) 
show that strong mass loss takes
place. Since we know the stellar and planet parameters for these cases
we can also calculate the mass loss through the Roche Lobe point. The
formula in Eq.~\ref{eq:general} would be then
\begin{equation}\label{eq:formularoche}
\dot M_{\rm XUV} \sim \frac{3 F_{\rm XUV}}{4\, {\rm G}\,K\,\rho}\, .
\end{equation}
where the $K$ parameter is defined as a function of $R_{\rm P}$,
$M_{\rm P}$ (planet 
radius and mass), $a_{\rm P}$ (semimajor axis) and $M_*$ (stellar mass), as
described in \citet{erk07}. Table~\ref{tab:roche} list the parameters
used in 
the calculation. Note that \citet{erk07} calculates the wrong values
of $K$, likely because of a mistake in the substitution of the units
used to determine
the Roche lobe distance, overestimating
the mass losses through the Roche lobe.
Fig.~\ref{masslosshistory} represents the mass loss history,
assuming that planet density remains constant and only thermal losses
are eroding the atmosphere. 
We consider also the hypothesis of mass losses following the
same trend (constant radius for HD~209458~b, HD~189733~b and
2M1207~A~b, $R_{\rm P}\sim 2.45~M_{\rm P}$ for GJ~436)
found in the mass-radius relation of close-in planets
(Eq.~\ref{eq:massradius}), in better 
agreement with theoretical models.
Future development of Eq.~\ref{eq:formularoche} 
should give a more accurate view
of the thermal losses. In this sense Lammer et al. \citep[in prep.,
  see also][]{lam09}  multiply
Eq.~\ref{eq:formularoche} by 
the heating efficiency (about 10--25\%). This would imply much
lower effects of evaporation, but it is insufficient to justify the
mass loss rate measured in HD~209458~b (see below).
Other models include different effects of EUV radiation in the
atmosphere, and infrared cooling \citep{yel04,tia05,gar07,mur09},
usually 
yielding smaller escape rates than Eq.~\ref{eq:formularoche}. Those
models are not easy to test with our sample, and they do not
consider the effects of X-rays.

%vvvvvvvvvvvvvvvvvvvvvvvvvvvvvvvvvvvvvvvvvvvvvvvvvvvvv
\section{Discussion}
The distribution of planets with the X-ray flux received seems to indicate
that planets have lost mass in their first stages.
Fig.~\ref{masses} can be interpreted in the same manner as an HR
diagram: the lack of planets in a given area of the diagram indicates
that they spend little time in that phase, while the accumulation of
planets in other area indicates that they spend much time in that
position. We can divide the diagram in four boxes based on the mass (at
1.5~M$_{\rm J}$) and the X-ray flux at the planet orbit (at $\log
F_{\rm X}$=2.15). We see then that only 1 out of 12 of the planets
with high flux have a high mass, while 38 out of 84 (45\%) of the
planets receiving lower fluxes have a high mass (up to 9~M$_{\rm J}$
in this diagram). Therefore it is clear
the absence of high mass planets suffering high flux levels. The
same exercise is applicable to Fig.~\ref{masses2}.

%----------------------------------  Fig. 7
%----------------------------------------------

To explain this distribution we propose three alternatives: 
({\em i}) an observational bias,
({\em ii}) an effect of planet formation, and ({\em iii}) a physical phenomenon
that moves the planets from their original positions in the diagram. The
observational biases \citep[discussed in][]{san10} easily explain the
lack of planets in the lower left corner of the diagram, corresponding
to low mass planets at long distances from the star. But the same biases
should make us find more planets in the upper right corner, where no
planets are found.  
The second possibility, the effects of planet formation, can not be
easily identified at the present level of knowledge. Simulations
carried out by, e.g., \citet{mor09} do not cover planets with 
$a_{\rm P}< 0.1$~a.u, and more recent simulations
\citep{ben11} might justify only the gap observed in
%Fig.~\ref{distances} 
the mass-distance diagram
at $M\la 1$\,M$_{\rm J}$ and $a_{\rm P}\la 0.04$\,a.u.  
The effects of
X-rays in an eventual photoevaporation of the protoplanetary disc has been
analyzed by \citet{dra09} and \citet{owen11}.
If planet 
formation favors presence of low mass planets close to the star, 
Figs.~\ref{masses} and \ref{masses2} just reflect a lack of high mass 
population at short
distances ($F_{\rm X} \propto d^{-2}$). 
However the distribution of density with mass at short
distance of the star \citep{san10} reveals that such population
exists (32\% of the 120 planets currently in the sample have $M > 1.5$\,M$_{\rm J}$), but they have an
increasing density with mass, an important detail also for
({\em iii}). Finally there exists the possibility that one or several 
physical processes related to the XUV emission (and X-rays as a proxy)
are eroding the atmospheres of planets at close 
distances to the star, yielding in the long term an uneven
distribution of masses with $F_{\rm X}$. The Eq.~\ref{eq:massloss}
indicates also that planet density provides protection against thermal
losses, consistent with the observed trend in the mass-density
  diagram just mentioned.
The distribution of masses with the XUV flux
accumulated over time (Fig.~\ref{agemasses}) further supports the
interpretation of mass losses as the effect of XUV irradiation, either by
thermal or non-thermal effects: most massive exoplanets that were 
initially exposed to high radiation would have now less than
$1.5$\,M$_{\rm J}$. 

%----------------------------------  Table 7
\begin{table}
\caption[]{Mass lost in planets with known density}\label{tab:roche}
\tabcolsep 2.1pt
\begin{center}
%\begin{scriptsize}
\begin{tabular}{lcccccccc}
\hline \hline
%--------------------------------------------------------------
{Planet name} & $R_{\rm P}$  & 
{$M_{\rm P}$}  & $\rho$ & $M_*$ & $a_{\rm P}$ & 
$K$\tablefootmark{a} & \multicolumn{2}{c}{$M_{\rm lost}$\tablefootmark{b}} (M$_{\rm J}$)\\
 & (R$_{\rm J}$) & (M$_{\rm J}$) & (g\,cm$^{-3}$) & (M$_\odot$) &
(a.u.) & & {(1)} & {(2)} \\
\hline
%--------------------------------------------------------------
GJ 436 b   & 0.365 & 0.0737 & 1.88 & 0.452 & 0.0289 & 0.76 & {0.06} & {0.08} \\
HD 189733 b& 1.151 & 1.15   & 0.94 & 0.8   & 0.0314 & 0.67 & {0.11} & {0.11} \\
HD 209458 b& 1.38  & 0.714  & 0.34 & 1.0   & 0.0475 & 0.65 & {0.18} & {0.17} \\
2M1207 A b & 1.5   & 4.0    & 1.47 & 0.025 & 46     & 1.00 & 2e-8 & {2e-8}\\
%--------------------------------------------------------------
\hline
\end{tabular}
%\end{scriptsize}
\end{center}
\vspace{-3mm}
\tablefoot{
\tablefoottext{a}{$K$ parameter to account for the Roche Lobe effects
  as in Eq.~\ref{eq:formularoche}.}
\tablefoottext{b}{Mass lost since 20~Myr old,
  including Roche lobe effects and thermal losses. (1) is for constant density, (2) with $M-R$ relation as explained in the text}}
\end{table}
%}
%---------------------------------------------

The upper limit of the hydrogen mass loss rate we calculate for
HD~209458\,b ($1\times10^{10}$ g\,s$^{-1}$) 
is consistent with the values of $\sim10^{10}$~g\,s$^{-1}$
\citep{vid03} interpreted by the authors
as hydrogen escaping the planet atmosphere, and with the calculation
of $\sim 8\times10^{10}$~g\,s$^{-1}$ 
extrapolated from \ion{C}{ii} line absorption by
\citet{lin10}. Similarly our calculation for HD~189733
($2\times10^{11}$ g\,s$^{-1}$) is consistent with the value of
$\sim10^{11}$~g\,s$^{-1}$ estimated by \citet{lec10}. 
Further support for our
interpretation is found if we look at the chromosphere: 
\citet{har10} finds that there is a correlation between chromospheric
activity and planet surface gravity. Contrary to the statement by
\citet{sch10}, 
this relation supports our conclusions, since denser planets also have
a stronger surface gravity. Therefore the accumulated effects of
erosion over time favor a resulting distribution with
denser (and with higher surface gravity) planets close
to active stars.

The presence of an ``erosion line'' (Fig.~\ref{masses})
that might separate the stages of strong and weak evaporation
  \citep{san10}, cannot be precisely quantified until 
we have a large sample of planets with known density and X-ray
measurements of their parent stars. It is possible that such a line
separates the phase of heavy erosion of a planet from that of slow
or no erosion. The establishment of such a line would be interesting for
future works to test whether a planet is still suffering strong
erosion. Planets above the line are potential targets to detect atmospheric
features.

%----------------------------------  Fig. 8
%----------------------------------------------

%vvvvvvvvvvvvvvvvvvvvvvvvvvvvvvvvvvvvvvvvvvvvvvvvvvvvv
\section{Conclusions}
The bulk of evidence support the effects of erosion on planet
atmospheres, as an effect of XUV radiation. The accumulated
effects of evaporation by radiation yield a population of only low
mass planets exposed to a currently high XUV radiation. This
interpretation is also supported by the lack of low density massive
planets among the close-in planets population.
We have used a simple physical
model to test the observed distribution. This model assumes that mass
loses are controlled by the thermal evaporation due to the XUV
radiation absorbed in the atmosphere. In the four planets of the
sample with known density we have reconstructed the mass loss
history, starting at a stellar age of 20~Myr, including thermal losses and
losses through the Roche lobe. The density protects the planet
atmosphere from further losses, but in a low density planet such as HD
209458b, up to $\sim$0.2~M$_{\rm J}$ 
have been lost according to this model.
Future model developments should 
take into account non-thermal losses, the role of planetary magnetic
field, and impact of different atmospheric composition. This research
will benefit from the future inclusion of a larger population of
planets with known density and X-ray stellar emission.

\begin{acknowledgements}
JSF and DGA acknowledge support from the
Spanish MICINN through grant AYA2008-02038 
and the Ram\'on y Cajal Program ref. RYC-2005-000549. 
IR acknowledges support from the Spanish MICINN via grant
AYA2006-15623-C02-01. 
GM acknowledges financial contribution from PRIN/INAF 
(P.I.: Lanza). This research has made use of the NASA's High Energy
Astrophysics Science Archive Research Center (HEASARC) 
and the public archives of XMM-Newton and Chandra.
We are grateful to the anonymous referee and to the
editor, T. Guillot, for the careful reading of and useful comments
on the manuscript.
\end{acknowledgements}

%--------------------------------------------------------------------
\Online

%vvvvvvvvvvvvvvvvvvvvvvvvvvvvvvvvvvvvvvvvvvvvvvvvvvvvv
%vvvvvvvvvvvvvvvvvvvvvvvvvvvvvvvvvvvvvvvvvvvvvvvvvvvvv
\begin{appendix}

% A. vvvvvvvvvvvvvvvvvvvvvvvvvvvvvvvvvvvvvvvvvvvvvvvvvvvvv
\section{Extrapolation of the lower temperature EMD}\label{loweremd}
The determination of the EMD in the transition region ($\log T$\,[K]$\sim
4.2-5.8$) usually benefits from the information provided by UV
lines. For the sources with no UV spectroscopic observations we need
to develop a method to calculate the EMD in this region.
We extrapolate the values of the EMD at
those temperatures based on the coronal counterpart, for which a
general proportionality seems to be present. Both transition region
and coronal material are supposed to be part of the same geometrical
structures (loops). In the coronal EMD of all sources we can identify 
the presence of material at $\log T$\,[K]$\sim
6.3$, the typical temperature of the solar corona, despite of their
activity level. We use the EM level at that peak, averaged over three
values of T, to calibrate the relation to the lower temperature
EMD. 
We used a sample of objects with a well calculated EMD over the whole
range, using same technique in all cases
\citep{san02,sanadleo,san03b,hue03,san04}, and adding $\alpha$~Cen~B
(Sect.~\ref{sec:highres}). We separated the sample 
in three groups depending on the level of activity (interpreted from
the amount of EMD found at the highest temperatures): 
low activity
stars (group 1: Procyon, $\alpha$~Cen~B), moderately active stars
(group 2: $\epsilon$~Eri, $\xi$~UMa~B), and active stars
(group 3: VY~Ari, $\sigma^2$~CrB, AR~Lac, FK~Aqr, AD~Leo, UX~Ari,
V711~Tau, II~Peg, AB~Dor). 

\begin{figure}
  \centering
  \includegraphics[width=0.49\textwidth]{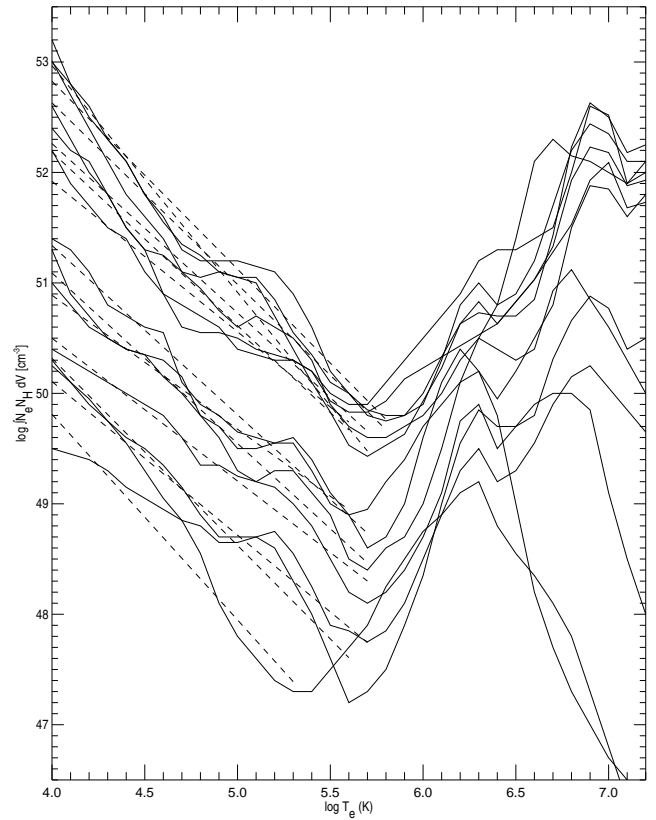}
  \caption{Linear fits (dashed lines) applied to the cool side of the
    EMD (solid lines) of well
    known coronal models.}\label{fig:TRemd}
\end{figure}

The lower temperature EMD can be defined using three parameters
(see Fig.~\ref{fig:TRemd}, Table~\ref{tab:fittremds}). 
Two come from the fitting of the EMD with a straight line: the 
slope of this line and the difference between the minimum EM (at $T_{\rm
  min}$) and the local maximum at  $\log T$\,[K]=$6.2-6.4$ 
($\Delta$EM$_1$). 
The fit makes use of values in the temperature range
$\log T$\,[K]=$4.2-T_{\rm min}$. Since the groups 1 and 2 have only 4
objects in total we applied only one fit to all of them.

We need also a way to account for the different sampling of the EMD in
T, from the 0.1 dex binning used in the EMD to the 3-temperature fit typical
in low resolution spectra. The fits with 1 or 2 temperature are
assumed to be like the 3-T fits with the remaining temperatures
considered as negligible. The third parameter needed in our model
accounts for this binning, in the form of a vertical shift of the EM
($\Delta$EM$_2$) to be added to $\Delta$EM$_1$. This parameter shows a
dependence on the level of activity, according to the distribution of
mass in temperature.
We used a representative star for each
group, all of them of spectral type K2V: 
$\alpha$~Cen~B (group 1), $\epsilon$~Eri (group 2), and AB~Dor
(group 3). The $\Delta$EM$_2$ of each case is listed in
Table~\ref{tab:fittremds}.

%----------------------------------  Fig. a1
%----------------------------------------------

Depending on the temperature and EM found in the targets in our sample
we use one of the three groups, and extrapolate the EM of the
transition region using the value of EM
at the temperature closer to $\log T$\,(K)=6.3 (EM$_{6.3}$): 
we first determine the
EM of $T_{\rm min}$ (using $\log T_{\rm min}$\,(K)=$5.7$): 
EM$_{\rm min}$=EM$_{\log T\sim 6.3} - \Delta$EM$_1 - \Delta$EM$_2$. Then we
extend the EM at lower temperatures with a straight line with the
slope in Table~\ref{tab:fittremds}, resulting in the values listed in
Table~\ref{tab:coolemd}. Uncertainties in the 
lower temperature EMD are calculated based on those from 
Table~\ref{tab:fittremds}. 

We made some tests to check the accuracy of the calculation with this
method. We used the same three representative stars ($\alpha$~Cen B,
$\epsilon$~Eri, and AB Dor) with a complete EMD
calculated using UV lines, and compared to the flux in same spectral
ranges using 3-T model combined with the extrapolated EM at lower
temperatures. The values measured from both models
(Table~\ref{tab:compmodels}) are very similar, 
so we are confident that the approach followed is correct.

Finally we compared the calculation of the EUV flux of $\epsilon$~Eri
with the direct EUVE spectrum. The luminosity in the band $80-170$~\AA\ in
the observed spectrum was 3.2e+27 erg\,s$^{-1}$, in the model based on
the whole EMD was 2.7e+27, and in the model based on the
3T+extrapolated EMs we get 1.9e+27.  These differences are very small
considering the process followed to get the synthetic spectra. We are
confident that the method can be safely applied to all late type stars
(late F to mid M spectral types). 
%\end{appendix}

%----------------------------------  Table A1
\begin{table}
\caption[]{Transition region EMD. Fit parameters}\label{tab:fittremds}
%\tabcolsep 3.pt
\begin{center}
%\begin{scriptsize}
\begin{tabular}{lccc}
\hline \hline
%--------------------------------------------------------------
{Activity level} & Slope  & $\Delta$EM$_1$ (cm$^{-3}$)& $\Delta$EM$_2$
(cm$^{-3}$) \\
\hline
%--------------------------------------------------------------
Low          & $-1.66\pm 0.15$  & $1.84\pm 0.36$ & 0.6 \\
Medium       & $-1.66\pm 0.15$  & $1.84\pm 0.36$ & 0.1 \\ 
High         & $-1.53\pm 0.26$  & $1.19\pm 0.25$ & 1.8 \\ 
%--------------------------------------------------------------
\hline
\end{tabular}
%\end{scriptsize}
\end{center}
\vspace{-3mm}
\end{table}
%}
%---------------------------------------------

\input{tablecoolemds}

%----------------------------------  Table A3
\begin{table}
\caption[]{Comparison of fluxes depending on models
  used\tablefootmark{a}}\label{tab:compmodels} 
\tabcolsep 2.8pt
\begin{center}
%\begin{scriptsize}
\begin{tabular}{c|cc|cc|cc}
\hline \hline
%--------------------------------------------------------------
  & \multicolumn{6}{c}{$L$ (erg\,s$^{-1}$)} \\
Range  & \multicolumn{2}{c}{$\alpha$~Cen~B}  & 
\multicolumn{2}{c}{$\epsilon$~Eri} & \multicolumn{2}{c}{AB Dor} \\
(\AA) & EMD & 3T & EMD & 3T & EMD & 3T \\
\hline
%--------------------------------------------------------------
$5-100$   & 7.0e+26 & 9.3e+26 & 1.8e+28 & 1.6e+28 & 2.1e+30 & 1.6e+30  \\
$100-920$ & 6.8e+27 & 4.1e+27 & 2.7e+28 & 1.7e+28 & 6.3e+29 & 4.2e+29  \\

%--------------------------------------------------------------
\hline
\end{tabular}
%\end{scriptsize}
\end{center}
\vspace{-3mm}
\tablefoot{
\tablefoottext{a}{The ``3T'' model includes also the predicted EM
  values in the transition region}}
\end{table}
%}
%---------------------------------------------

%----------------------------------
%\begin{appendix}
% B. vvvvvvvvvvvvvvvvvvvvvvvvvvvvvvvvvvvvvvvvvvvvvvvvvvvvv
\section{Emission Measure Distribution of $\alpha$~Cen B}\label{sec:highres}
\begin{figure}[b]
   \centering
   \includegraphics[angle=270,width=0.49\textwidth]{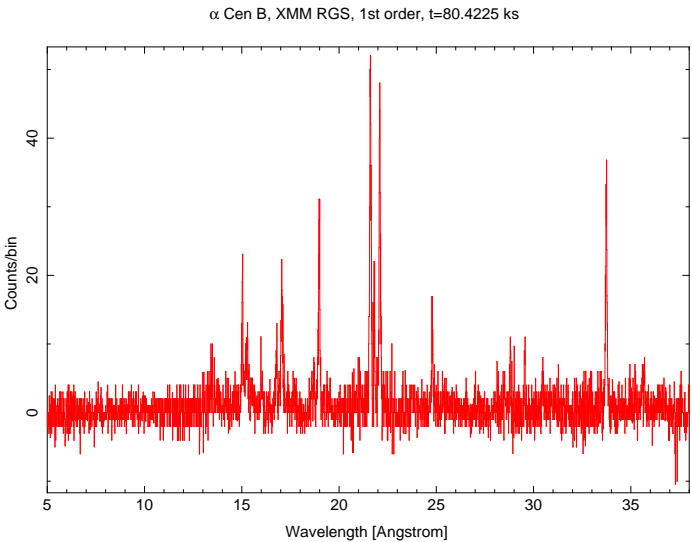}
   \caption{XMM-Newton RGS combined spectrum of $\alpha$~Cen B}\label{specacenb}
   \end{figure}

\begin{figure}
   \centering
   \includegraphics[width=0.49\textwidth]{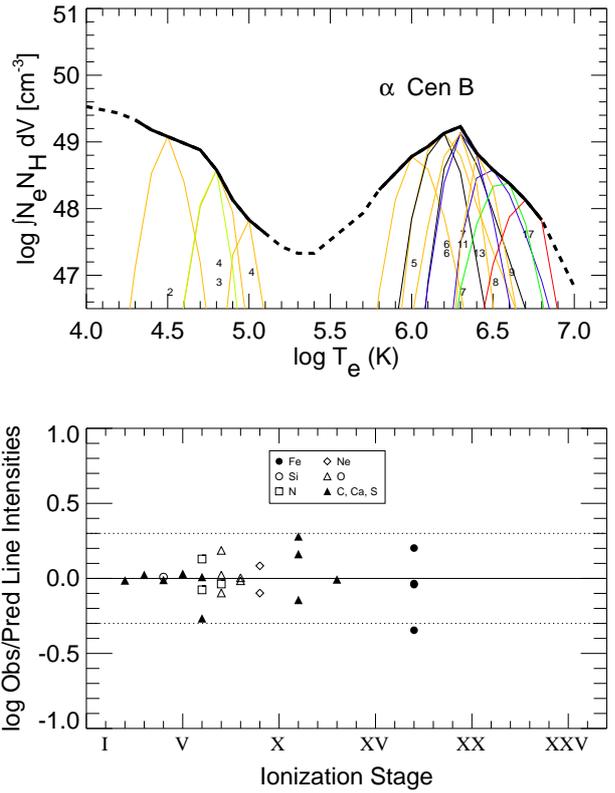}
   \caption{Upper panel: Emission Measure Distribution of $\alpha$~Cen B. 
     Thin lines represent the relative contribution function for each
     ion (the emissivity function multiplied by the EMD at each
     point). Small numbers indicate the ionization stages of the
     species. Lower panel indicates the observed-to-predicted line
     flux ratios for the ion stages in the upper figure. The dotted
     lines denote a factor of 2.}\label{emdacenb}
   \end{figure}

We have calculated the Emission Measure Distribution (EMD) of the K2V star
$\alpha$~Cen~B, needed to test the extrapolation of the lower EMD
temperature and the synthesis of the EUV spectra. We use the UV lines fluxes
measured by \citet{san03b}, and the XMM-Newton/RGS lines fluxes listed
in Table~\ref{tab:acenfluxes}, from an observation taken on Jan
2009 (Fig.~\ref{specacenb}). The coronal model (the EMD) was
constructed following 
\citet{san03a}. The resulting EMD (Table~\ref{tab:emd}) is displayed
in Fig.~\ref{emdacenb}, with coronal abundances as listed in
Table~\ref{tab:abundances}. A global fit to the Chandra/LETG spectrum
was applied by \citet{raa03}, with similar results in the corona.

\input{fluxesacen}

%------------------------------------------------- begin table
%  Emission Measure Distribution  Table B.2
\begin{table}
\caption{Emission measure distribution of $\alpha$~Cen~B}\label{tab:emd}
%\tabcolsep 3.pt
\begin{center}
\begin{small}
\begin{tabular}{cc}
\hline \hline
{log~$T$ (K)} & {EM 
  (cm$^{-3}$)\tablefootmark{a}} \\
\hline
%----------------
4.0 & 49.53:  \\
4.1 & 49.48:  \\
4.2 & 49.43:  \\
4.3 & 49.33$^{+0.20}_{-0.30}$  \\
4.4 & 49.18$^{+0.10}_{-0.20}$  \\
4.5 & 49.08$^{+0.10}_{-0.20}$  \\
4.6 & 48.98$^{+0.20}_{-0.20}$  \\
4.7 & 48.88$^{+0.10}_{-0.30}$  \\
4.8 & 48.58$^{+0.20}_{-0.20}$  \\
4.9 & 48.13$^{+0.10}_{-0.10}$  \\
5.0 & 47.83$^{+0.10}_{-0.30}$  \\
5.1 & 47.63$^{+0.20}_{-0.30}$  \\
5.2 & 47.43:  \\
5.3 & 47.33:  \\
5.4 & 47.33:  \\
5.5 & 47.53:  \\
5.6 & 47.73:  \\
5.7 & 47.93:  \\
5.8 & 48.28$^{+0.40}_{-0.40}$  \\
5.9 & 48.53$^{+0.30}_{-0.40}$  \\
6.0 & 48.78$^{+0.20}_{-0.30}$  \\
6.1 & 48.93$^{+0.10}_{-0.40}$  \\
6.2 & 49.13$^{+0.20}_{-0.20}$  \\
6.3 & 49.23$^{+0.10}_{-0.20}$  \\
6.4 & 48.83$^{+0.10}_{-0.20}$  \\
6.5 & 48.58$^{+0.10}_{-0.30}$  \\
6.6 & 48.38$^{+0.20}_{-0.20}$  \\
6.7 & 48.13$^{+0.20}_{-0.40}$  \\
6.8 & 47.83$^{+0.10}_{-0.30}$  \\
6.9 & 47.33:  \\
7.0 & 46.83:  \\
%---------------
\hline
\end{tabular}
\end{small}
\end{center}
\tablefoot{
\tablefoottext{a}{Emission measure (EM=log $\int N_{\rm e} N_{\rm H}
  {\rm d}V$), where $N_{\rm e}$ 
and $N_{\rm H}$ are electron and hydrogen densities, in
cm$^{-3}$. Error bars provided are not independent
between the different temperatures, as explained in \citet{san03b}}}.
\end{table}
%--------------------------------------------- end table

%----------------------------------  Fig. B1
%----------------------------------------------

%----------------------------------  Fig. B2
%----------------------------------------------

%------------------------------------------------- begin table
%  Element abundances  Table B.3
\begin{table}
\caption{Coronal abundances of $\alpha$~Cen~B (solar
  units\tablefootmark{a}).}\label{tab:abundances} 
%\tabcolsep 3.pt
\begin{center}
\begin{footnotesize}
%\begin{tabular}{lccccccccc}
 \begin{tabular}{lrcccr}
\hline \hline
{X} & {FIP (eV)} & Ref.$^a$ & (AG89$^a$) & {[X/H]} \\
\hline
%----------------
  C & 11.26 &  8.39 & (8.56) &  0.04$\pm$ 0.11 \\
  N & 14.53 &  7.78 & (8.05) & -0.01$\pm$ 0.14 \\
  O & 13.61 &  8.66 & (8.93) & -0.40$\pm$ 0.12 \\
 Ne & 21.56 &  7.84 & (8.09) & -0.48$\pm$ 0.19 \\
 Si &  8.15 &  7.51 & (7.55) & -0.44$\pm$ 0.10 \\
  S & 10.36 &  7.14 & (7.21) &  0.49$\pm$ 0.18 \\
 Ca &  6.11 &  6.31 & (6.36) &  0.22$\pm$ 0.27 \\
 Fe &  7.87 &  7.45 & (7.67) &  0.19$\pm$ 0.19 \\
%----------------
\hline
\end{tabular}
\end{footnotesize}
\end{center}
\tablefoot{
\tablefoottext{a}{Solar photospheric abundances from \citet{asp05},
  adopted in this table, are expressed in logarithmic scale. 
Note that several values have been
updated in the literature since \citet[AG89]{anders}, also listed in
parenthesis for easier comparison.}}
\end{table}
%--------------------------------------------- end table

% C. vvvvvvvvvvvvvvvvvvvvvvvvvvvvvvvvvvvvvvvvvvvvvvvvvvvvv
\section{The X-Exoplanets Data Server}\label{app:xexo}

The X-Exoplanets data
server\footnote{http://sdc.cab.inta-csic.es/xexoplanets/} provides
information on the planet-bearing stars that have been observed with
XMM-Newton or Chandra. In the near future, synthetic spectra covering
the EUV range \citep{san09} and EUVE
data will also be available. The system contains reduced,
science-ready data and was set up to facilitate analysis of the
effects of coronal radiation on exoplanets atmospheres.

\subsection{Functionalities: Search}

The X-Exoplanets data server is accessed by means of a web-based
fill-in form that permits queries by list of objects and coordinates
and radius. Searches can be customized to include physical parameters
of the stars and planets as well as light curves and reduced spectra
obtained from XMM-Newton and Chandra data.

\subsection{Functionalities: Results}

\begin{figure}
   \centering
   \includegraphics[width=0.49\textwidth]{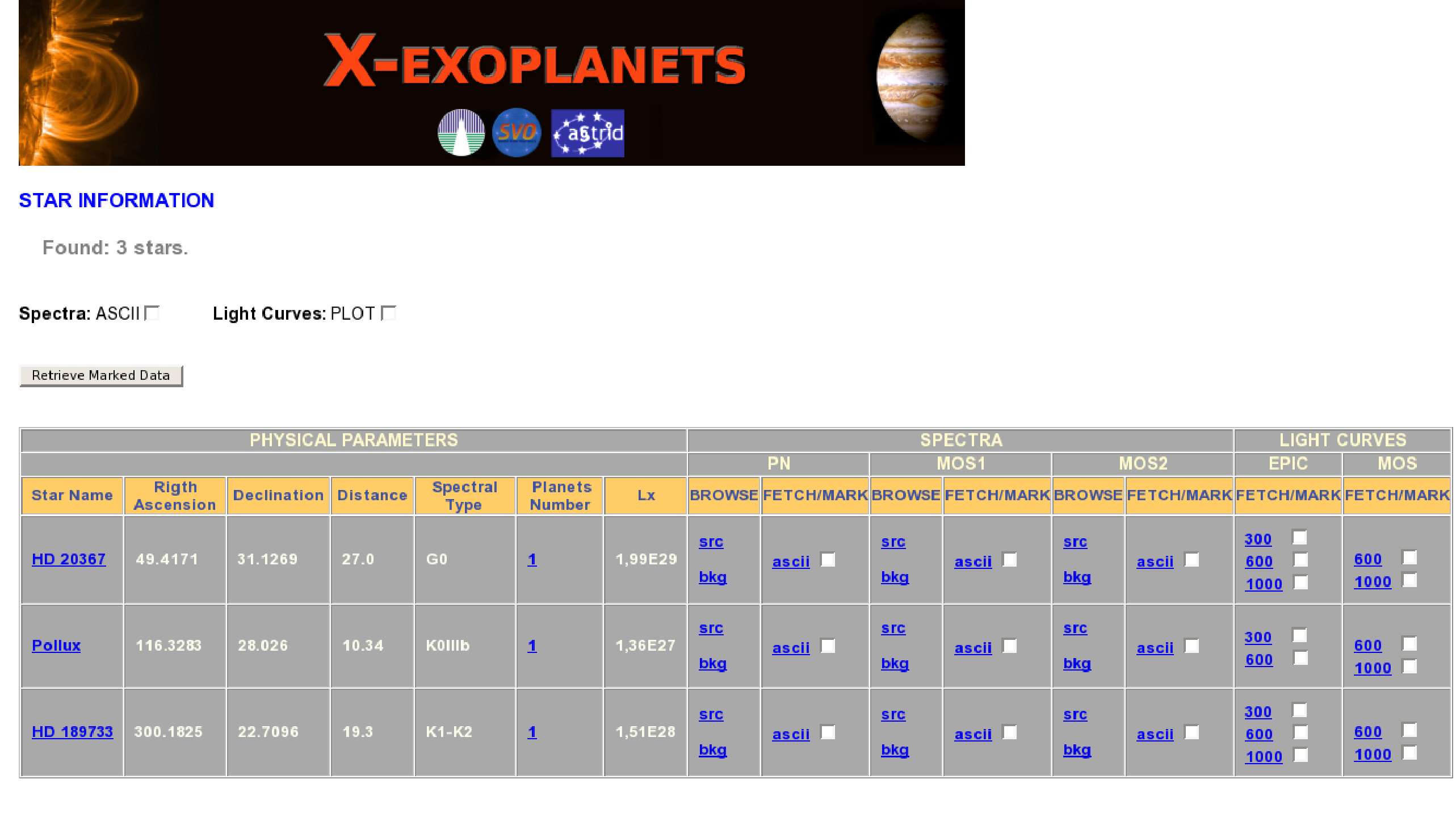}
   \caption{The X-Exoplanet Data Server. Result of a query}\label{pantalla}
   \end{figure}

\begin{figure*}
   \centering
   \includegraphics[angle=90,width=0.49\textwidth]{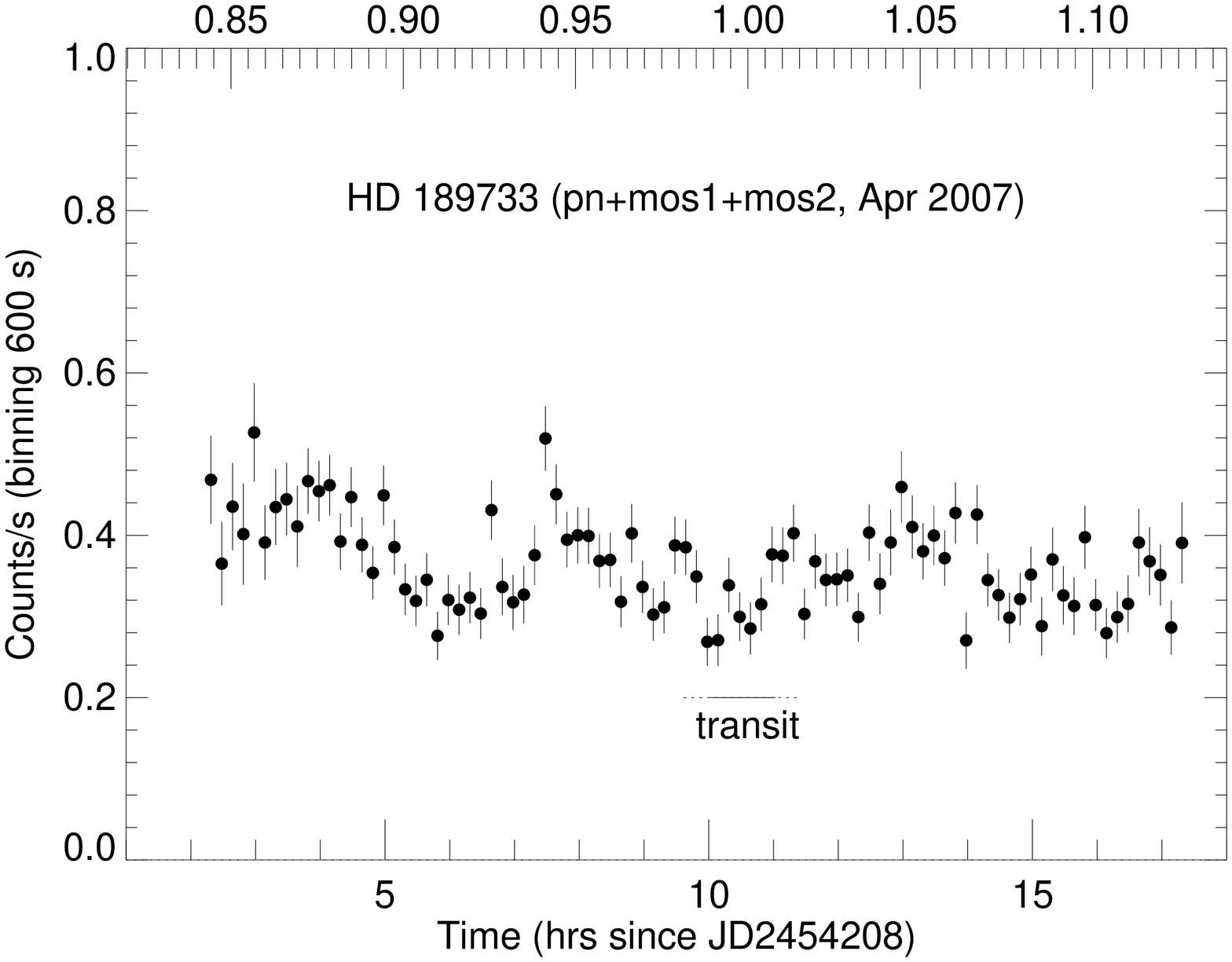}
   \includegraphics[width=0.44\textwidth]{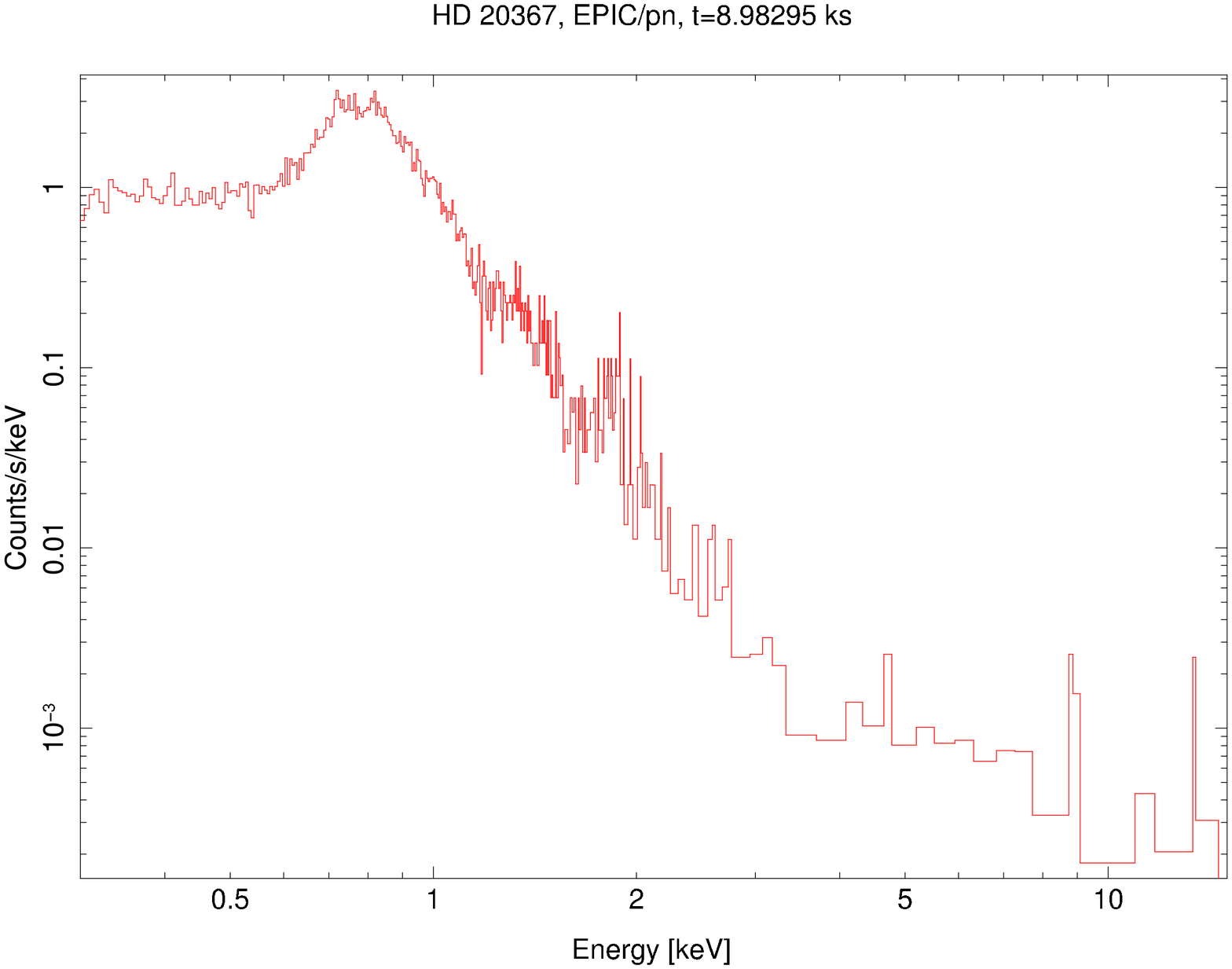}
   \caption{The X-Exoplanet Data Server. Light curve (left) and reduced
  spectrum (right). In the light curve of HD~189733 we mark the
  orbital phase \citep{win07} of HD~189733~b in the upper axis, as well as the
  interval when the transit takes place (partial in dotted line, total in solid
  line).}\label{xexo:examples}
   \end{figure*}

An example of the result of a query is given in
Fig.~\ref{pantalla}. Light curves and reduced spectra can be
visualized by clicking on the corresponding link
(Fig.~\ref{xexo:examples}).The system incorporates multidownload and 
preview capabilities. Links to SIMBAD and the Extrasolar Planet
Encyclopaedia are also provided. 

\subsection{The Virtual Observatory service}

VO-compliance of an astronomical archive constitutes an added value of
enormous importance for the optimum scientific exploitation of their
datasets. The X-Exoplanet service has been designed following the IVOA
standards and requirements. In particular, it implements the SSA (Simple
Spectral Access) protocol and its associated data model, a standard
defined for retrieving 1-D data.

%----------------------------------  Fig. C1
%----------------------------------------------

%----------------------------------  Fig. C2
%----------------------------------------------

\end{appendix}

\end{document}

%% file: tabobslog.tex
%----------------------------------  Table 1
%\onltab{1}{
\begin{table}
\caption[]{Observation log of stars with exoplanets$^a$}\label{tabobslog}
\tabcolsep 3.pt
\begin{center}
\begin{scriptsize}
  \begin{tabular}{lrrcccrc}
\hline \hline
{Star name} & \multicolumn{2}{c}{Measured coordinates} & {Date} & {Instr.\tablefootmark{a}} & {t} & {S/N} & Notes\\ 
{} & \multicolumn{2}{c}{$\alpha$, $\delta$ (J2000.0)} & & & (ks) & & \\ 
\hline
%--------------------------------------------------------------
14 Her              & 16:10:24.6 & $+43$:49:01 & 2005/09/11 & EPIC &   5 &    4.9 &  \\
16 Cyg B            & 19:41:48.9 & $+50$:31:28 & 2008/11/08 & EPIC &  11 &    2.0 &  \\
2M1207 A            & 12:07:33.5 & $-39$:32:54 & 2003/03/03 & ACIS &  50 &    0.4 &  \\
30 Ari B            & 02:36:57.6 & $+24$:38:52 & 2001/01/16 & EPIC &  20 &  244   & \\
 & & & 2001/01/17 & EPIC &  34 &  & \\
47 UMa              & 10:59:28.4 & $+40$:25:46 & 2006/06/11 & EPIC &   8 &    4.8 &  \\
51 Peg              & 22:57:28.1 & $+20$:46:08 & 2008/12/06 & ACIS &   5 &    2.6 &  \\
 55 Cnc             & 08:52:35.7 & $+28$:19:47 & 2009/04/11 & EPIC &  12 &   14.0 &  \\
$\beta$ Pic         & 05:47:17.1 & $-51$:03:59 & 2004/01/04 & EPIC &  68 &    5.7 &  \\
$\epsilon$ Eri      & 03:32:55.9 & $-09$:27:31 & 2003/01/19 & EPIC &  12 &  297   &  \\
 GJ 86              & 02:10:28.1 & $-50$:49:19 & 2008/06/10 & EPIC &  15 &   43.0 & WD in field  \\
 GJ 317             & 08:40:59.0 & $-23$:27:15 & 2009/04/20 & EPIC &  18 &    9.6 &  \\
GJ 436              & 11:42:11.6 & $+26$:42:16 & 2008/12/10 & EPIC &  30 &   14.5 &  \\
 GJ 674             & 17:28:40.3 & $-46$:53:50 & 2008/09/05 & EPIC &  44 &  179   &  \\
 GJ 876             & 22:53:17.3 & $-14$:15:55 & 2008/11/14 & EPIC &  23 &   34.7 &  \\
GQ Lup              & 15:49:12.1 & $-35$:39:05 & 2008/08/16 & EPIC &   8 &   31.7 &  \\
 HD 4308            & 00:44:39.4 & $-65$:39:05 & 2008/12/02 & EPIC &   9 &    2.5 &  \\
HD 20367            & 03:17:40.1 & $+31$:07:37 & 2005/02/11 & EPIC &  10 &  140   &  \\
 HD 27442           & 04:16:29.0 & $-59$:18:09 & 2009/02/10 & EPIC &   7 &    2.7 &  \\
HD 46375            & 06:33:12.4 & $+05$:27:49 & 2005/10/14 & EPIC &   8 &    7.3 &  \\
HD 49674            & 06:51:30.9 & $+40$:52:03 & 2006/04/10 & EPIC &   8 &    6.5 &  \\
HD 50554            & 06:54:42.8 & $+24$:14:43 & 2006/04/16 & EPIC &   9 &    2.7 &  \\
 HD 52265           & 07:00:18.0 & $-05$:22:01 & 2008/09/19 & EPIC &   9 &    5.4 &  \\
HD 70642            & 08:21:28.2 & $-39$:42:18 & 2006/04/08 & EPIC &  13 &    4.2 &  \\
HD 75289            & 08:47:40.1 & $-41$:44:14 & 2005/04/28 & EPIC &   8 &    2.3 &  \\
 HD 93083           & 10:44:20.9 & $-33$:34:38 & 2008/05/26 & EPIC &  12 &    6.3 &  \\
HD 95089            & 10:58:47.7 & $+01$:43:44 & 2009/05/26 & EPIC &  37 &    0.8 &  \\
 HD 99492           & 11:26:45.9 & $+03$:00:24 & 2008/06/19 & EPIC &  24 &   11.3 &  \\
HD 101930           & 11:43:30.1 & $-58$:00:21 & 2009/01/06 & EPIC &   2 &    0.4 &  \\
 HD 102195          & 11:45:42.2 & $+02$:49:16 & 2008/06/15 & EPIC &  18 &   53.1 &  \\
HD 108147           & 12:25:46.2 & $-64$:01:20 & 2002/08/10 & EPIC &   6 &    4.2 &  \\
 HD 111232          & 12:48:51.8 & $-68$:25:29 & 2008/07/29 & EPIC &   9 &    0.9 &  \\
 HD 114386          & 13:10:39.7 & $-35$:03:20 & 2008/07/29 & EPIC &   9 &    3.0 &  \\
HD 114762           & 13:12:19.7 & $+17$:31:02 & 2004/06/28 & EPIC &  29 &    1.4 & dM in field \\
 HD 114783          & 13:12:43.7 & $-02$:15:54 & 2009/01/22 & EPIC &   8 &    3.0 &  \\
HD 130322           & 14:47:32.8 & $-00$:16:54 & 2005/07/21 & EPIC &   7 &    7.7 &  \\
 HD 154345          & 17:02:36.5 & $+47$:05:02 & 2008/12/25 & EPIC &   8 &    9.0 &  \\
 HD 164922          & 18:02:33.4 & $+26$:18:43 & 2009/03/19 & EPIC &   9 &    1.0 &  \\
HD 179949           & 19:15:33.3 & $-24$:10:46 & 2005/05/21 & ACIS &  30 &  101   & \\
 & & & 2005/05/22 & ACIS &  30 &  & \\
 & & & 2005/05/29 & ACIS &  30 &  & \\
 & & & 2005/05/30 & ACIS &  32 &  & \\
 & & & 2005/05/31 & ACIS &  30 &  & \\
HD 187123           & 19:46:57.9 & $+34$:25:09 & 2006/04/21 & EPIC &  16 &    1.4 &  \\
HD 189733           & 20:00:43.8 & $+22$:42:34 & 2007/04/17 & EPIC &  43 &   92.5 &  \\
HD 190360           & 20:03:37.9 & $+29$:53:45 & 2005/04/25 & EPIC &   4 &    1.4 &  \\
HD 195019           & 20:28:18.6 & $+18$:46:10 & 2006/04/24 & EPIC &  10 &    2.7 & dK in field \\
HD 209458           & 22:03:10.8 & $+18$:53:03 & 2006/11/15 & EPIC &  31 &    1.8 &  \\
HD 216435           & 22:53:38.1 & $-48$:35:55 & 2006/04/21 & EPIC &   7 &   11.9 &  \\
HD 216437           & 22:54:39.6 & $-70$:04:26 & 2005/04/13 & EPIC &   6 &    4.0 &  \\
HD 217107           & 22:58:15.7 & $-02$:23:43 & 2005/05/16 & EPIC &   7 &    2.3 &  \\
HD 218566           & 23:09:10.9 & $-02$:15:39 & 2001/06/10 & EPIC &   3 &    6.8 & \\
 & & & 2004/06/05 & EPIC &  10 &  & \\
HD 330075           & 15:49:37.7 & $-49$:57:48 & 2005/08/07 & EPIC &  16 &    3.1 &  \\
HR 8799             & 23:07:28.8 & $+21$:08:02 & 2009/08/30 & ACIS &  10 &    6.5 &  \\
 $\mu$ Ara          & 17:44:08.7 & $-51$:50:04 & 2008/09/06 & EPIC &   6 &    2.9 &  \\
 & & & 2008/10/02 & EPIC &   9 &  &  \\
NGC 2423 3          & 07:37:09.2 & $-13$:54:24 & 2008/05/05 & EPIC &   9 &    1.1 &  \\
Pollux              & 07:45:18.8 & $+28$:01:33 & 2001/04/26 & EPIC &  32 &   34.0 &  \\
$\tau$ Boo          & 13:47:15.9 & $+17$:27:22 & 2003/06/24 & EPIC &  56 &  317   & dM2 in field \\
$\upsilon$ And      & 01:36:47.7 & $+41$:24:15 & 2009/10/20 & ACIS &  15 &   52.0 & \\
 & & & 2009/10/22 & ACIS &  15 & &  \\
 & & & 2009/10/27 & ACIS &  14 & &  \\
 & & & 2009/10/29 & ACIS &  14 & &  \\
%---------------------------------------------
\hline
\end{tabular}
\end{scriptsize}
\end{center}
\vspace{-3mm}
 \tablefoot{
 \tablefoottext{a}{XMM-Newton (EPIC) or Chandra (ACIS) instrument used to measure the X-ray flux.}}
\end{table}
%}
%---------------------------------------------

%% file: tabrosatlog.tex
%----------------------------------  Table 2
%\onltab{2}{
\begin{table}
\caption[]{ROSAT/PSPC X-ray flux ($0.12-2.48$~keV) of stars with exoplanets}\label{tabrosatlog}
\tabcolsep 3.pt
\begin{center}
\begin{scriptsize}
  \begin{tabular}{lrrccl}
\hline \hline
%--------------------------------------------------------------
{Star name} & \multicolumn{2}{c}{Coordinates} & S/N & {$\log f_{\rm X}$} & {Notes} \\
{} & \multicolumn{2}{c}{$\alpha$, $\delta$ (J2000.0)} & & (erg s$^{-1}$cm$^{-2}$) & \\ 
\hline
%--------------------------------------------------------------
18 Del              & 20:58:25.9 & $+10$:50:21 &    4.5 & $ -12.48$ &  \\
1RXS 1609           & 16:09:30.3 & $-21$:04:58 &    3.5 & $ -12.52$ &  \\
4 UMa               & 08:40:12.8 & $+64$:19:40 &   11.6 & $ -13.26$ &  K-M in field \\
61 Vir              & 13:18:24.3 & $-18$:18:40 &    5.7 & $ -13.06$ &  \\
$\alpha$ Ari        & 02:07:10.4 & $+23$:27:44 &    5.1 & $ -13.50$ &  \\
BD-10 3166          & 10:58:28.8 & $-10$:46:13 &    3.1 & $<-12.52$ &  +dM5, uncertain $d$ \\
$\gamma$ Cep        & 23:39:20.8 & $+77$:37:56 &    8.1 & $<-13.03$ &  dM4 in field \\
GJ 176              & 04:42:55.8 & $+18$:57:29 &    3.2 & $ -12.62$ &  \\
GJ 832              & 21:33:34.0 & $-49$:00:32 &    7.3 & $ -12.69$ &  \\
GJ 3021             & 00:16:12.7 & $-79$:51:04 &    7.8 & $<-11.62$ &  dM4 in field \\
HD 3651             & 00:39:21.8 & $+21$:15:01 &    4.5 & $ -12.89$ &  \\
HD 10647            & 01:42:29.3 & $-53$:44:27 &    N/A & $ -12.24$ &  \\
HD 38529            & 05:46:34.9 & $+01$:10:05 &    5.3 & $ -12.37$ &  \\
HD 41004 A          & 05:59:49.6 & $-48$:14:22 &    7.5 & $<-12.02$ &  dM2 in field \\
HD 48265            & 06:40:01.7 & $-48$:32:31 &    6.0 & $ -12.43$ &  \\
HD 70573            & 08:22:50.0 & $+01$:51:33 &    4.0 & $ -12.31$ &  uncertain $d$ \\
HD 87883            & 10:08:43.1 & $+34$:14:32 &    N/A & $ -12.99$ &  \\
HD 89744            & 10:22:10.6 & $+41$:13:46 &    7.6 & $ -13.17$ &  \\
HD 102365           & 11:46:31.1 & $-40$:30:01 &    4.8 & $ -13.20$ &  faint dM4 in field\\
HD 128311           & 14:36:00.6 & $+09$:44:47 &    7.5 & $ -12.04$ &  \\
HD 142415           & 15:57:40.8 & $-60$:12:00 &    5.0 & $ -12.50$ &  \\
HD 147513           & 16:24:01.3 & $-39$:11:34 &   16.2 & $ -11.40$ &  \\
HD 150706           & 16:31:17.6 & $+79$:47:23 &   12.0 & $ -12.13$ &  \\
HD 169830           & 18:27:49.5 & $-29$:49:00 &   16.8 & $ -12.94$ &  \\
HD 176051           & 18:57:01.5 & $+32$:54:06 &    7.2 & $<-12.27$ &  F9V in field \\
HIP 75458           & 15:24:55.8 & $+58$:57:57 &    3.5 & $ -13.66$ &  \\
HIP 79431           & 16:12:41.8 & $-18$:52:31 &    3.6 & $ -13.50$ &  ROSAT/WGA detector \\
HR 810              & 02:42:33.5 & $-50$:48:01 &    7.0 & $ -11.67$ &  \\
%---------------------------------------------
\hline
\end{tabular}
\end{scriptsize}
\end{center}
\vspace{-3mm}
\tablefoot{The star GJ 667C (M1V) was well detected with ROSAT/PSPC (S/N=7.1), but its X-ray emission is attributed to its companions GJ 667A (K3V) and GJ 667B (K5V), at 43\arcsec.}
\end{table}
%}
%---------------------------------------------

%% file: tabfits.tex
%----------------------------------  Table 2
\onltab{3}{
\begin{table*}
\caption[]{X-ray flux ($0.12-2.48$~keV) and fits of stars with exoplanets (XMM-Newton and Chandra data)}\label{tabfits}
\tabcolsep 4.pt
\begin{center}
\begin{scriptsize}
  \begin{tabular}{lcccccc}
\hline \hline
{Star name} & {$\log f_{\rm X}$} & {log T} & {log EM} & {Elements (X)} &  {[X/H]} \\ 
{} & (erg s$^{-1}$cm$^{-2}$) & (K) & (cm$^{-3}$) & & \citep[Solar values of][]{anders}\\ 
\hline
%--------------------------------------------------------------
14 Her              & $ -13.67$ &$6.49^{+0.06}_{-0.06}$ & $49.58^{+0.10}_{-0.12}$ & Fe & $-0.22^{+0.00}_{-0.00}$ \\
 16 Cyg B           & $<-14.04$ &$6.30$ & $49.40^{+0.24}_{-0.57}$ & Fe & $-0.20$ \\
2M1207 A            & $<-15.27$ &$6.30$ & $48.52^{+0.34}_{-0.00}$ & Fe & $-0.20$ \\
30 Ari B            & $ -11.74$ &$6.66^{+0.01}_{-0.01}$, $6.30^{+0.08}_{-0.06}$, $6.91^{+0.01}_{-0.00}$ & $52.02^{+0.03}_{-0.03}$, $51.05^{+0.17}_{-0.13}$, $51.61^{+0.03}_{-0.03}$ & Fe, O, Ne & $-0.01^{+0.02}_{-0.02}$, $-0.48^{+0.04}_{-0.04}$, $-0.17^{+0.06}_{-0.07}$ \\
47 UMa              & $ -14.92$ &$6.20^{+0.16}_{-0.20}$ & $48.23^{+0.24}_{-0.45}$ & Fe & $-0.10^{+0.00}_{-0.20}$ \\
51 Peg              & $<-13.93$ &$6.30$ & $49.22^{+0.24}_{-0.59}$ & Fe & $-0.20$ \\
 55 Cnc             & $ -13.62$ &$6.65^{+0.05}_{-0.09}$ & $49.35^{+0.06}_{-0.11}$ & Fe & $-0.30^{+0.20}_{-0.00}$ \\
$\beta$ Pic         & $ -15.01$ &$6.29^{+0.10}_{-0.06}$ & $48.35^{+0.11}_{-0.19}$ & Fe & $-0.30^{+0.20}_{-0.00}$ \\
$\epsilon$ Eri      & $ -10.89$ &$6.60^{+0.01}_{-0.01}$, $6.27^{+0.05}_{-0.06}$, $6.87^{+0.02}_{-0.02}$ & $50.67^{+0.05}_{-0.08}$, $50.29^{+0.07}_{-0.09}$, $50.22^{+0.07}_{-0.08}$ & Fe, O, N, Ne,  C & $-0.17^{+0.03}_{-0.02}$, $-0.25^{+0.07}_{-0.05}$, $ 0.30^{+0.09}_{-0.07}$, $-0.14^{+0.05}_{-0.04}$, $ 0.25^{+0.17}_{-0.15}$ \\
 GJ 317             & $ -13.86$ &$6.62^{+0.09}_{-0.09}$ & $48.81^{+0.12}_{-0.17}$ & Fe & $-0.23^{+0.46}_{-0.44}$ \\
GJ 436              & $ -14.13$ &$6.76^{+0.15}_{-0.13}$, $6.26^{+0.19}_{-0.20}$ & $48.59^{+0.17}_{-0.32}$, $48.22^{+0.22}_{-0.55}$ & Fe & $-0.74^{+0.45}_{-1.69}$ \\
 GJ 674             & $ -11.67$ &$6.88^{+0.08}_{-0.05}$, $6.48^{+0.01}_{-0.01}$, $7.13^{+0.05}_{-0.02}$ & $49.62^{+0.12}_{-0.15}$, $50.44^{+0.03}_{-0.03}$, $49.71^{+0.04}_{-0.14}$ & Fe, O, Ne & $-0.27^{+0.05}_{-0.05}$, $-0.51^{+0.03}_{-0.03}$, $-0.13^{+0.07}_{-0.07}$ \\
 GJ 86              & $<-12.74$ &$6.54^{+0.02}_{-0.02}$ & $50.06^{+0.03}_{-0.03}$ & Fe & $-0.05^{+0.11}_{-0.11}$ \\
 GJ 876             & $ -13.26$ &$6.65^{+0.02}_{-0.02}$, $6.18^{+0.00}_{-0.00}$ & $48.84^{+0.02}_{-0.03}$, $47.93^{+0.18}_{-0.33}$ & Fe & $-0.37^{+0.06}_{-0.07}$ \\
GQ Lup              & $ -12.92$ &$6.89^{+0.06}_{-0.07}$, $7.60^{+0.06}_{-0.07}$ & $52.43^{+0.06}_{-0.24}$, $52.71^{+0.04}_{-0.06}$ & Fe & $-0.30^{+0.09}_{-0.00}$ \\
 HD 4308            & $<-14.82$ &$6.30$ & $48.64^{+0.28}_{-1.12}$ & Fe & $-0.20$ \\
HD 20367            & $ -11.65$ &$6.30^{+0.99}_{-0.11}$, $6.65^{+0.12}_{-0.21}$, $6.95^{+0.03}_{-0.02}$ & $50.71^{+0.98}_{-0.44}$, $51.79^{+0.05}_{-0.60}$, $51.39^{+0.05}_{-0.07}$ & Fe, O, N, Ne & $-0.01^{+0.04}_{-0.05}$, $-0.36^{+0.10}_{-0.07}$, $ 0.19^{+0.18}_{-0.20}$, $-0.47^{+0.21}_{-0.40}$ \\
 HD 27442           & $<-14.46$ &$6.30$ & $48.83^{+0.24}_{-0.61}$ & Fe & $-0.20$ \\
HD 46375            & $ -13.96$ &$6.63^{+0.12}_{-0.10}$ & $49.87^{+0.09}_{-0.17}$ & Fe & $-0.30^{+0.00}_{-0.00}$ \\
HD 49674            & $ -13.89$ &$6.49^{+0.18}_{-0.15}$, $6.00^{+0.00}_{-0.00}$ & $50.13^{+0.11}_{-0.97}$, $50.83^{+0.00}_{-0.00}$ & Fe,  O & $ 0.41^{+0.37}_{-0.55}$, $-0.41^{+0.19}_{-0.86}$ \\
HD 50554            & $<-14.74$ &$6.30$ & $49.02^{+0.22}_{-0.46}$ & Fe & $-0.20$ \\
 HD 52265           & $ -14.08$ &$6.43^{+0.12}_{-0.07}$ & $49.46^{+0.23}_{-0.33}$ & Fe & $ 0.27^{+0.51}_{-N/A}$ \\
HD 70642            & $ -14.61$ &$6.53^{+0.21}_{-0.12}$ & $49.08^{+0.16}_{-0.29}$ & Fe & $-0.30^{+0.20}_{-0.00}$ \\
HD 75289            & $<-15.10$ &$6.30$ & $48.60^{+0.34}_{-N/A}$ & Fe & $-0.20$ \\
 HD 93083           & $ -14.10$ &$6.77^{+0.17}_{-0.22}$ & $49.74^{+0.15}_{-0.26}$ & Fe & $-1.12^{+0.98}_{-N/A}$ \\
HD 95089            & $<-15.18$ &$6.30$ & $47.94^{+0.31}_{-N/A}$ & Fe & $-0.20$ \\
 HD 99492           & $ -14.04$ &$6.59^{+0.10}_{-0.07}$ & $49.23^{+0.12}_{-0.15}$ & Fe & $-0.22^{+0.42}_{-0.59}$ \\
HD 101930           & $ -14.99$ &$6.30$ & $48.75^{+0.41}_{-N/A}$ & Fe & $-0.20$ \\
 HD 102195          & $ -12.57$ &$6.74^{+0.02}_{-0.02}$, $6.01^{+0.04}_{-0.01}$ & $50.80^{+0.04}_{-0.05}$, $51.04^{+0.08}_{-0.14}$ & Fe & $-0.05^{+0.07}_{-0.06}$ \\
HD 108147           & $ -13.86$ &$6.60^{+0.16}_{-0.13}$ & $50.05^{+0.15}_{-0.26}$ & Fe & $-0.10^{+0.00}_{-0.00}$ \\
 HD 111232          & $<-14.65$ &$6.30$ & $49.05^{+0.06}_{-0.00}$ & Fe & $-0.20$ \\
 HD 114386          & $ -14.44$ &$6.39^{+0.20}_{-0.22}$ & $49.20^{+0.20}_{-0.39}$ & Fe & $-0.30^{+0.20}_{-0.00}$ \\
HD 114762           & $<-14.78$ &$6.30$ & $49.19^{+0.43}_{-N/A}$ & Fe & $-0.20$ \\
 HD 114783          & $<-14.16$ &$6.38^{+0.12}_{-0.09}$ & $49.21^{+0.17}_{-0.33}$ & Fe & $-0.30^{+0.20}_{-0.00}$ \\
HD 130322           & $ -13.78$ &$6.56^{+0.09}_{-0.08}$ & $49.94^{+0.10}_{-0.14}$ & Fe & $-0.24^{+0.14}_{-0.06}$ \\
 HD 154345          & $ -13.47$ &$6.42^{+0.08}_{-0.05}$ & $49.67^{+0.11}_{-0.14}$ & Fe & $ 0.31^{+0.40}_{-0.78}$ \\
 HD 164922          & $<-15.02$ &$6.30$ & $48.44^{+0.41}_{-N/A}$ & Fe & $-0.20$ \\
HD 179949           & $ -12.56$ &$6.06^{+0.05}_{-0.04}$, $6.72^{+0.01}_{-0.01}$ & $50.77^{+0.15}_{-0.17}$, $50.74^{+0.03}_{-0.04}$ & Fe & $ 0.12^{+0.04}_{-0.04}$ \\
HD 187123           & $<-14.20$ &$6.30$ & $49.98^{+0.01}_{-0.00}$ & Fe & $-0.20$ \\
HD 189733           & $ -12.47$ &$6.85^{+0.01}_{-0.01}$, $6.13^{+0.01}_{-0.01}$, $6.68^{+0.04}_{-0.05}$ & $50.66^{+0.02}_{-0.05}$, $50.49^{+0.03}_{-0.04}$, $50.23^{+0.04}_{-0.14}$ & Fe, O, Ne & $-0.40^{+0.00}_{-0.00}$, $-0.05^{+0.02}_{-0.04}$, $-0.40^{+0.00}_{-0.00}$ \\
HD 190360           & $<-14.13$ &$6.30$ & $49.05^{+0.01}_{-0.00}$ & Fe & $-0.20$ \\
HD 195019           & $<-14.99$ &$6.30$ & $48.91^{+0.26}_{-0.74}$ & Fe & $ 0.00^{+0.00}_{-0.00}$ \\
HD 209458           & $<-15.02$ &$6.30$ & $49.10^{+0.06}_{-0.00}$ & Fe & $-0.20$ \\
HD 216435           & $ -13.38$ &$6.58^{+0.04}_{-0.03}$ & $50.36^{+0.05}_{-0.06}$ & Fe & $-0.00^{+0.00}_{-0.23}$ \\
HD 216437           & $ -14.30$ &$6.40^{+0.21}_{-0.10}$ & $49.28^{+0.19}_{-0.36}$ & Fe & $-0.30^{+0.20}_{-0.00}$ \\
HD 217107           & $<-15.17$ &$6.30$ & $48.20^{+0.45}_{-N/A}$ & Fe & $-0.20$ \\
HD 218566           & $ -13.99$ &$6.52^{+0.17}_{-0.16}$ & $49.72^{+0.21}_{-0.39}$ & Fe & $-0.24^{+0.92}_{-N/A}$ \\
HD 330075           & $ -14.97$ &$6.60^{+1.30}_{-0.24}$ & $49.22^{+0.30}_{-N/A}$ & Fe & $-0.30^{+0.20}_{-0.00}$ \\
HR 8799             & $ -13.26$ &$6.54^{+0.07}_{-0.08}$ & $50.75^{+0.11}_{-0.14}$ & Fe & $-0.55^{+0.60}_{-N/A}$ \\
 $\mu$ Ara          & $<-14.46$ &$6.30$ & $48.68^{+0.16}_{-0.26}$ & Fe & $-0.20$ \\
NGC 2423 3          & $<-14.58$ &$6.30$ & $56.27^{+1.58}_{-N/A}$ & Fe & $-0.20$ \\
Pollux              & $ -12.97$ &$6.23^{+0.05}_{-0.12}$, $6.54^{+0.21}_{-0.43}$ & $49.90^{+0.06}_{-0.11}$, $48.70^{+1.08}_{-1.68}$ & Fe,  O & $ 0.43^{+0.35}_{-0.34}$, $-0.36^{+0.30}_{-0.15}$ \\
$\tau$ Boo          & $ -11.51$ &$6.62^{+0.01}_{-0.01}$, $6.30^{+0.04}_{-0.04}$, $6.92^{+0.01}_{-0.01}$ & $51.70^{+0.02}_{-0.02}$, $50.94^{+0.08}_{-0.05}$, $50.93^{+0.06}_{-0.04}$ & Fe, O, Ne, Mg & $-0.28^{+0.00}_{-0.00}$, $-0.56^{+0.03}_{-0.02}$, $-0.51^{+0.05}_{-0.06}$, $-0.28^{+0.05}_{-0.03}$ \\
                    & & & & Si, C,  N & $-0.09^{+0.05}_{-0.04}$, $-0.22^{+0.00}_{-0.00}$, $-0.37^{+0.00}_{-0.00}$ \\
$\upsilon$ And      & $ -12.77$ &$6.53^{+0.02}_{-0.02}$ & $50.09^{+0.03}_{-0.04}$ & Fe & $ 0.30^{+0.09}_{-0.09}$ \\
%---------------------------------------------
\hline
\end{tabular}
\end{scriptsize}
\end{center}
\vspace{-3mm}
% \scriptsize{Notes
%}
\end{table*}
}
%---------------------------------------------

%% file: tabbands.tex
%----------------------------------  Table with fluxes in different bands
\onltab{4}{
\begin{table*}
\caption[]{Stars with exoplanets. XUV luminosity predicted in different bands\tablefootmark{a}}\label{tab:bands}
%\tabcolsep 3.pt
\begin{center}
%\begin{scriptsize}
  \begin{tabular}{lcccccc}
\hline \hline
Star & \multicolumn{6}{c}{Luminosity (erg\,s$^{-1}$) in the wavelength ranges indicated (in \AA)} \\
     & 100--200 & 200--300 & 300--400 & 400--550 & 550--700 & 700--920 \\
\hline
%--------------------------------------------------------------
\\
14 Her         & 26.01$\pm$0.08 & 26.61$\pm$0.15 & 27.10$\pm$0.39 & 26.63$\pm$0.39 & 27.06$\pm$0.44 & 27.34$\pm$0.48 \\
 16 Cyg B      & $<$26.32 & $<$27.00 & $<$27.37 & $<$26.89 & $<$27.33 & $<$27.65 \\
2M1207 A       & $<$25.44 & $<$26.12 & $<$26.49 & $<$26.01 & $<$26.45 & $<$26.77 \\
30 Ari B       & 28.65$\pm$0.00 & 28.90$\pm$0.00 & 28.83$\pm$0.05 & 28.19$\pm$0.05 & 27.92$\pm$0.20 & 28.02$\pm$0.35 \\
47 UMa         & 25.64$\pm$0.01 & 25.88$\pm$0.04 & 25.86$\pm$0.31 & 25.29$\pm$0.38 & 25.74$\pm$0.42 & 26.00$\pm$0.48 \\
51 Peg         & $<$26.14 & $<$26.82 & $<$27.19 & $<$26.71 & $<$27.15 & $<$27.47 \\
 55 Cnc        & 25.68$\pm$0.09 & 26.10$\pm$0.28 & 26.84$\pm$0.43 & 26.37$\pm$0.42 & 26.83$\pm$0.44 & 27.11$\pm$0.49 \\
$\beta$ Pic    & 25.31$\pm$0.07 & 25.91$\pm$0.14 & 26.34$\pm$0.43 & 25.87$\pm$0.42 & 26.33$\pm$0.44 & 26.61$\pm$0.49 \\
$\epsilon$ Eri & 27.69$\pm$0.00 & 27.96$\pm$0.00 & 27.81$\pm$0.00 & 27.26$\pm$0.00 & 27.31$\pm$0.00 & 27.49$\pm$0.00 \\
 GJ 86         & $<$26.85 & $<$27.64 & $<$28.50 & $<$28.02 & $<$28.49 & $<$28.81 \\
 GJ 317        & 25.32$\pm$0.19 & 25.95$\pm$0.36 & 26.77$\pm$0.45 & 26.31$\pm$0.45 & 26.79$\pm$0.44 & 27.07$\pm$0.49 \\
GJ 436         & 25.26$\pm$0.05 & 25.66$\pm$0.19 & 26.21$\pm$0.42 & 25.75$\pm$0.41 & 26.21$\pm$0.44 & 26.48$\pm$0.48 \\
 GJ 674        & 26.94$\pm$0.01 & 27.37$\pm$0.03 & 27.49$\pm$0.24 & 26.97$\pm$0.20 & 27.02$\pm$0.37 & 27.29$\pm$0.46 \\
 GJ 876        & 25.44$\pm$0.02 & 25.65$\pm$0.10 & 26.01$\pm$0.34 & 25.52$\pm$0.35 & 25.93$\pm$0.42 & 26.20$\pm$0.48 \\
GQ Lup         & 29.17$\pm$0.11 & 29.51$\pm$0.41 & 30.38$\pm$0.47 & 29.92$\pm$0.45 & 30.41$\pm$0.44 & 30.69$\pm$0.49 \\
 HD 4308       & $<$25.56 & $<$26.24 & $<$26.61 & $<$26.13 & $<$26.57 & $<$26.89 \\
HD 20367       & 28.44$\pm$0.01 & 28.67$\pm$0.05 & 28.90$\pm$0.26 & 28.25$\pm$0.26 & 28.51$\pm$0.42 & 28.81$\pm$0.48 \\
 HD 27442      & $<$25.75 & $<$26.43 & $<$26.80 & $<$26.32 & $<$26.76 & $<$27.08 \\
HD 46375       & 26.37$\pm$0.18 & 27.00$\pm$0.36 & 27.83$\pm$0.46 & 27.37$\pm$0.45 & 27.85$\pm$0.44 & 28.13$\pm$0.49 \\
HD 49674       & 29.06$\pm$0.00 & 28.43$\pm$0.04 & 28.61$\pm$0.21 & 27.88$\pm$0.32 & 28.11$\pm$0.38 & 28.41$\pm$0.44 \\
HD 50554       & $<$25.94 & $<$26.62 & $<$26.99 & $<$26.51 & $<$26.95 & $<$27.27 \\
 HD 52265      & 26.05$\pm$0.08 & 26.97$\pm$0.05 & 27.11$\pm$0.29 & 26.54$\pm$0.36 & 26.95$\pm$0.44 & 27.22$\pm$0.48 \\
HD 70642       & 25.62$\pm$0.17 & 26.28$\pm$0.31 & 27.05$\pm$0.45 & 26.59$\pm$0.44 & 27.06$\pm$0.44 & 27.34$\pm$0.49 \\
HD 75289       & $<$25.52 & $<$26.20 & $<$26.57 & $<$26.09 & $<$26.53 & $<$26.85 \\
 HD 93083      & 26.14$\pm$0.19 & 26.81$\pm$0.42 & 27.69$\pm$0.47 & 27.23$\pm$0.45 & 27.72$\pm$0.44 & 28.00$\pm$0.49 \\
HD 95089       & $<$26.71 & $<$27.39 & $<$27.76 & $<$27.28 & $<$27.72 & $<$28.04 \\
 HD 99492      & 25.75$\pm$0.18 & 26.39$\pm$0.34 & 27.20$\pm$0.45 & 26.73$\pm$0.44 & 27.21$\pm$0.44 & 27.49$\pm$0.49 \\
HD 101930      & 25.63$\pm$0.03 & 26.27$\pm$0.05 & 26.32$\pm$0.35 & 25.84$\pm$0.36 & 26.24$\pm$0.43 & 26.52$\pm$0.48 \\
 HD 102195     & 28.84$\pm$0.01 & 28.51$\pm$0.17 & 29.09$\pm$0.36 & 28.57$\pm$0.41 & 29.03$\pm$0.43 & 29.31$\pm$0.48 \\
HD 108147      & 26.59$\pm$0.19 & 27.21$\pm$0.34 & 28.02$\pm$0.45 & 27.55$\pm$0.44 & 28.03$\pm$0.44 & 28.31$\pm$0.49 \\
 HD 111232     & $<$25.97 & $<$26.65 & $<$27.02 & $<$26.54 & $<$26.98 & $<$27.30 \\
 HD 114386     & 25.79$\pm$0.05 & 26.45$\pm$0.10 & 26.73$\pm$0.38 & 26.27$\pm$0.37 & 26.69$\pm$0.44 & 26.96$\pm$0.48 \\
HD 114762      & $<$26.11 & $<$26.79 & $<$27.16 & $<$26.68 & $<$27.12 & $<$27.44 \\
 HD 114783     & $<$25.90 & $<$26.62 & $<$27.17 & $<$26.69 & $<$27.14 & $<$27.46 \\
HD 130322      & 26.47$\pm$0.18 & 27.13$\pm$0.32 & 27.91$\pm$0.45 & 27.44$\pm$0.44 & 27.92$\pm$0.44 & 28.20$\pm$0.49 \\
 HD 154345     & 26.29$\pm$0.08 & 27.24$\pm$0.05 & 27.34$\pm$0.28 & 26.76$\pm$0.36 & 27.16$\pm$0.44 & 27.43$\pm$0.48 \\
 HD 164922     & $<$25.36 & $<$26.04 & $<$26.41 & $<$25.93 & $<$26.37 & $<$26.69 \\
HD 179949      & 28.70$\pm$0.01 & 28.42$\pm$0.12 & 28.83$\pm$0.36 & 28.29$\pm$0.42 & 28.76$\pm$0.43 & 29.04$\pm$0.48 \\
HD 187123      & $<$26.90 & $<$27.58 & $<$27.95 & $<$27.47 & $<$27.91 & $<$28.23 \\
HD 189733      & 27.90$\pm$0.00 & 27.86$\pm$0.02 & 27.84$\pm$0.13 & 27.08$\pm$0.19 & 27.48$\pm$0.27 & 27.60$\pm$0.43 \\
HD 190360      & $<$25.97 & $<$26.65 & $<$27.02 & $<$26.54 & $<$26.98 & $<$27.30 \\
HD 195019      & $<$25.92 & $<$26.64 & $<$26.89 & $<$26.40 & $<$26.84 & $<$27.16 \\
HD 209458      & $<$26.02 & $<$26.70 & $<$27.07 & $<$26.59 & $<$27.03 & $<$27.35 \\
HD 216435      & 26.92$\pm$0.19 & 27.56$\pm$0.31 & 28.34$\pm$0.44 & 27.86$\pm$0.44 & 28.34$\pm$0.44 & 28.62$\pm$0.49 \\
HD 216437      & 25.83$\pm$0.06 & 26.50$\pm$0.10 & 26.81$\pm$0.38 & 26.35$\pm$0.37 & 26.77$\pm$0.43 & 27.04$\pm$0.48 \\
HD 217107      & $<$25.12 & $<$25.80 & $<$26.17 & $<$25.69 & $<$26.13 & $<$26.45 \\
HD 218566      & 26.27$\pm$0.17 & 26.94$\pm$0.29 & 27.69$\pm$0.45 & 27.23$\pm$0.44 & 27.70$\pm$0.44 & 27.98$\pm$0.49 \\
HD 330075      & 25.56$\pm$0.09 & 26.02$\pm$0.25 & 26.71$\pm$0.42 & 26.25$\pm$0.41 & 26.70$\pm$0.44 & 26.98$\pm$0.49 \\
HR 8799        & 27.26$\pm$0.16 & 27.92$\pm$0.33 & 28.71$\pm$0.46 & 28.26$\pm$0.44 & 28.73$\pm$0.44 & 29.01$\pm$0.49 \\
$\mu$ Ara      & $<$25.60 & $<$26.28 & $<$26.65 & $<$26.17 & $<$26.61 & $<$26.93 \\
NGC 2423 3     & $<$33.19 & $<$33.87 & $<$34.24 & $<$33.76 & $<$34.20 & $<$34.52 \\
Pollux         & 27.68$\pm$0.01 & 27.98$\pm$0.01 & 27.66$\pm$0.23 & 26.93$\pm$0.34 & 27.18$\pm$0.41 & 27.47$\pm$0.48 \\
$\tau$ Boo     & 28.21$\pm$0.01 & 28.47$\pm$0.12 & 28.97$\pm$0.37 & 28.35$\pm$0.27 & 28.50$\pm$0.42 & 28.77$\pm$0.49 \\
$\upsilon$ And & 26.75$\pm$0.20 & 27.46$\pm$0.22 & 28.10$\pm$0.40 & 27.60$\pm$0.44 & 28.07$\pm$0.44 & 28.35$\pm$0.49 \\
\\
%---------------------------------------------
\hline
\end{tabular}
%\end{scriptsize}
\end{center}
\vspace{-3mm}
 \tablefoot{
 \tablefoottext{a}{Only stars with XMM-Newton or Chandra data.} 
}
\end{table*}
}
%---------------------------------------------

%% file: tabfluxrosat.tex
%----------------------------------  Table 2
%\onltab{1}{
\begin{table*}[t]
\caption[]{ROSAT X-ray (5--100 \AA) and EUV (100--920~\AA) luminosity of stars with exoplanets\tablefootmark{a} (ROSAT). XUV includes the 5--920~\AA\ range.}\label{tabfluxrosat}
\tabcolsep 3.pt
\begin{center}
\begin{scriptsize}
  \begin{tabular}{lcccccccccccccc}
\hline \hline
%\vspace{3mm}
%--------------------------------------------------------------
{Planet name} & {Sp. type} & {Stellar distance} & {$\log L_{\rm X}$} & {$\log L_{\rm EUV}$} &
{$\log L_{\rm bol}$} & age & $M_p \sin i$ & $a_p$ & $\log F_{\rm X}$ & $\log F_{\rm XUV}$ & 
$\log F_{\rm X}$accum. & $\log F_{\rm XUV}$accum. & $\rho \dot M_{\rm X}$ & $\rho \dot M_{\rm XUV}$  \\ 
{} & {(star)} & {(pc)} & {(erg s$^{-1}$)} & {(erg s$^{-1}$)} & {(erg s$^{-1}$)} & (Gyr) & {(m$_J$)} & {(a.u.)} & 
%{(erg s$^{-1}$cm$^{-2}$)} & {(erg cm$^{-2}$)} & (g$^2$s$^{-1}$cm$^{-3}$) 
\multicolumn{2}{c}{(erg s$^{-1}$cm$^{-2}$)} & \multicolumn{2}{c}{(erg cm$^{-2}$)} & 
\multicolumn{2}{c}{(g$^2$s$^{-1}$cm$^{-3}$)\tablefootmark{b}} \\ 
\hline
%--------------------------------------------------------------
%\multicolumn{11}{c}{ROSAT data}\\
18 Del b            &      G6III & $ 73.10\pm  3.74$ &     29.33 &     30.02 & 35.11 &   \dots  & 10.30 &   2.60 & $  1.05$ & $  1.82 $ & \dots &    \dots &      1.3e+08 &      7.5e+08 \\
1RXS1609 b          &        K7V & $145:$            &     29.88 &     30.50 & 33.20 &    0.09  &  8.00 & 330.00 & $ -2.61$ & $ -1.90 $ & 12.91 &    13.72 &      2.8e+04 &      1.4e+05 \\
4 UMa b             &      K1III & $ 77.40\pm  4.25$ &     28.41 &     29.23 & 35.65 &   \dots  &  7.10 &   0.87 & $  1.08$ & $  1.97 $ & \dots &    \dots &      1.4e+08 &      1.0e+09 \\
61 Vir b            &        G5V & $  8.53\pm  0.05$ &     26.88 &     27.92 & 33.49 &    7.96  &  0.02 &   0.05 & $  2.03$ & $  3.10 $ & 21.02 &    21.76 &      1.2e+09 &      1.4e+10 \\
61 Vir c            &            &                   &           &           &       &          &  0.06 &   0.22 & $  0.76$ & $  1.83$ & 19.75  &    21.76 &      6.4e+07 &      7.6e+08 \\
61 Vir d            &            &                   &           &           &       &          &  0.07 &   0.48 & $  0.08$ & $  1.15$ & 19.07  &    21.76 &      1.3e+07 &      1.6e+08 \\
$\alpha$ Ari b      &      K2III & $ 20.21\pm  0.40$ &     27.19 &     28.18 & 35.51 &   \dots  &  1.80 &   1.20 & $ -0.42$ & $  0.62 $ & \dots &    \dots &      4.3e+06 &      4.7e+07 \\
BD-10 3166 b        &        G4V & $ 66:           $ &  $<$29.20 &  $<$29.91 & 33.19 &    0.25  &  0.48 &   0.05 & $< 4.43$ & $< 5.21 $ & 20.84 &    21.61 &    (3.0e+11) &    (1.8e+12) \\
$\gamma$ Cep b      &       K2IV & $ 13.79\pm  0.10$ &  $<$27.33 &  $<$28.30 & 34.63 &    4.08: &  1.60 &   2.04 & $<-0.74$ & $< 0.28$ & 18.16: &   18.62: &    (2.0e+06) &    (2.1e+07) \\
GJ 176 b            &      M2.5V & $  9.42\pm  0.22$ &     27.41 &     28.37 & 31.74 &    3.62  &  0.03 &   0.07 & $  2.32$ & $  3.33 $ & 20.12 &    21.43 &      2.4e+09 &      2.4e+10 \\
GJ 832 b            &      M1.5V & $  4.94\pm  0.03$ &     26.78 &     27.83 & 31.81 &    9.24  &  0.64 &   3.40 & $ -1.73$ & $ -0.64 $ & 16.77 &    18.05 &      2.1e+05 &      2.5e+06 \\
GJ 3021 b           &        G6V & $ 17.62\pm  0.16$ &  $<$28.95 &  $<$29.70 & 33.39 &    0.37  &  3.37 &   0.49 & $< 2.12$ & $< 2.94 $ & 18.91 &    19.61 &    (1.5e+09) &    (9.8e+09) \\
HD 3651 b           &        K0V & $ 11.11\pm  0.09$ &     27.27 &     28.25 & 33.36 &    4.46  &  0.20 &   0.28 & $  0.91$ & $  1.94 $ & 19.47 &    20.23 &      9.2e+07 &      9.8e+08 \\
HD 10647 b          &        F8V & $ 17.35\pm  0.19$ &     28.31 &     29.15 & 33.75 &    0.95  &  0.93 &   2.03 & $  0.25$ & $  1.14 $ & 17.83 &    18.47 &      2.0e+07 &      1.6e+08 \\
HD 38529 b          &       G4IV & $ 42.43\pm  1.66$ &     28.96 &     29.71 & 34.41 &    0.36: &  0.78 &   0.13 & $  3.28$ & $  4.09$ & 20.42: &   20.86: &      2.1e+10 &      1.4e+11 \\
HD 39091 b          &       G1IV & $ 18.21\pm  0.15$ &     27.49 &     28.44 & 33.74 &    3.22: & 10.30 &   3.28 & $ -0.99$ & $  0.01$ & 17.44: &   18.12: &      1.1e+06 &      1.1e+07 \\
HD 41004 A b        &        K1V & $ 43.03\pm  1.89$ &  $<$29.31 &  $<$30.01 & 33.36 &    0.22  &  2.54 &   1.64 & $< 1.43$ & $< 2.21 $ & 17.79 &    18.50 &    (3.0e+08) &    (1.8e+09) \\
HD 48265 b          &        G5V & $ 87.41\pm  5.50$ &     29.53 &     30.20 & 34.18 &    0.16  &  1.16 &   1.51 & $  1.72$ & $  2.47 $ & 18.15 &    18.61 &      5.9e+08 &      3.3e+09 \\
HD 70573 b          &    G1-1.5V & $ 45.7:         $ &     29.09 &     29.82 & 33.33 &    0.30  &  6.10 &   1.76 & $  1.15$ & $  1.95 $ & 17.76 &    18.48 &      1.6e+08 &      1.0e+09 \\
HD 87883 b          &        K0V & $ 18.06\pm  0.31$ &     27.60 &     28.54 & 33.10 &    2.73  & 12.10 &   3.60 & $ -0.96$ & $  0.02 $ & 17.17 &    18.00 &      1.2e+06 &      1.2e+07 \\
HD 89744 b          &        F7V & $ 38.99\pm  1.06$ &     28.11 &     28.97 & 34.38 &    1.28  &  7.20 &   0.88 & $  0.77$ & $  1.69 $ & 18.79 &    19.28 &      6.6e+07 &      5.5e+08 \\
HD 102365 b         &        G2V & $  9.24\pm  0.06$ &     26.81 &     27.86 & 33.49 &    8.83  &  0.05 &   0.46 & $  0.04$ & $  1.12 $ & 19.10 &    19.84 &      1.2e+07 &      1.5e+08 \\
HD 128311 b         &        K0V & $ 16.57\pm  0.27$ &     28.48 &     29.29 & 33.06 &    0.74  &  2.18 &   1.10 & $  0.95$ & $  1.82 $ & 18.13 &    18.95 &      1.0e+08 &      7.5e+08 \\
HD 128311 c         &            &                   &           &           &       &          &  3.21 &   1.76 & $  0.54$ & $  1.41$ & 17.72  &    18.95 &      3.9e+07 &      2.9e+08 \\
HD 142415 b         &        G1V & $ 34.57\pm  1.00$ &     28.65 &     29.44 & 33.63 &    0.57  &  1.62 &   1.05 & $  1.16$ & $  2.01 $ & 18.35 &    19.01 &      1.6e+08 &      1.2e+09 \\
HD 147513 b         &     G3/G5V & $ 12.87\pm  0.14$ &     28.90 &     29.65 & 33.56 &    0.40  &  1.21 &   1.32 & $  1.21$ & $  2.03 $ & 18.11 &    18.77 &      1.8e+08 &      1.2e+09 \\
HD 150706 b         &         G0 & $ 27.23\pm  0.42$ &     28.82 &     29.59 & 33.55 &    0.45  &  1.00 &   0.82 & $  1.54$ & $  2.38 $ & 18.53 &    19.20 &      3.9e+08 &      2.7e+09 \\
HD 169830 b         &        F8V & $ 36.32\pm  1.20$ &     28.26 &     29.10 & 34.23 &    1.02  &  2.88 &   0.81 & $  0.99$ & $  1.90 $ & 18.80 &    19.32 &      1.1e+08 &      8.8e+08 \\
HD 169830 c         &            &                   &           &           &       &          &  4.04 &   3.60 & $ -0.30$ & $  0.60$ & 17.50  &    19.32 &      5.6e+06 &      4.5e+07 \\
HD 176051 b         &       K1V  & $ 14.98\pm  0.12$ &  $<$28.22 &  $<$29.07 & 32.86 &    1.09  &  1.50 &   1.76 & $< 0.28$ & $< 1.19 $ & 17.67 &    18.56 &    (2.1e+07) &    (1.7e+08) \\
HIP 75458 b         &      K2III & $ 31.33\pm  0.50$ &     27.41 &     28.37 & 35.38 &   \dots  &  8.82 &   1.27 & $ -0.25$ & $  0.76 $ & \dots &    \dots &      6.3e+06 &      6.4e+07 \\
HIP 79431 b         &        M3V & $ 14.90\pm  0.79$ &     26.89 &     27.93 & 60.02 &    7.84  &  2.10 &   0.36 & $  0.33$ & $  1.40 $ & 28.76 &    28.76 &      2.4e+07 &      2.8e+08 \\
HR 810 b            &  G0V       & $ 17.24\pm  0.16$ &     28.79 &     29.56 & 33.80 &    0.47  &  2.26 &   0.93 & $  1.41$ & $  2.25 $ & 18.49 &    19.11 &      2.9e+08 &      2.0e+09 \\
%---------------------------------------------
\hline
\end{tabular}
\end{scriptsize}
\end{center}
\vspace{-3mm}
 \tablefoot{
 \tablefoottext{a}{Planet data from The Extrasolar Planets Encyclopedia (http://exoplanet.eu).} 
 \tablefoottext{b}{1~M$_{\rm J}$\,Gyr$^{-1}$=$6.02 \times 10^{13}$~g\,s$^{-1}$}}
\end{table*}
%}
%---------------------------------------------

%% file: tabfluxes.tex
%----------------------------------  Table 2
%\onltab{1}{
\begin{table*}[th!]
\caption[]{X-ray (5--100 \AA) and EUV (100--920~\AA) luminosity of stars with exoplanets\tablefootmark{a} (XMM-Newton and Chandra).}\label{tabfluxes}
\tabcolsep 3.pt
\begin{center}
\begin{scriptsize}
  \begin{tabular}{lcccccccccccccc}
\hline \hline
{Planet name} & {Sp. type} & {Stellar distance} & {$\log L_{\rm X}$} & {$\log L_{\rm EUV}$}
& {$\log L_{\rm bol}$} & age & $M_p \sin i$ & $a_p$ & $\log F_{\rm X}$ & $\log F_{\rm XUV}$ & 
$\log F_{\rm X}$accum. & $\log F_{\rm XUV}$accum. & $\rho \dot M_{\rm X}$ & $\rho \dot M_{\rm XUV}$ \\ 
{} & {(star)} & {(pc)} & {(erg s$^{-1}$)} & {(erg s$^{-1}$)} & {(erg s$^{-1}$)} & (Gyr) & {(m$_J$)} & {(a.u.)} & 
%{(erg s$^{-1}$cm$^{-2}$)} & {(erg cm$^{-2}$)} & (g$^2$s$^{-1}$cm$^{-3}$) 
\multicolumn{2}{c}{(erg s$^{-1}$cm$^{-2}$)} & \multicolumn{2}{c}{(erg cm$^{-2}$)} & 
\multicolumn{2}{c}{(g$^2$s$^{-1}$cm$^{-3}$)\tablefootmark{c}}  \\ 
\hline
%--------------------------------------------------------------
14 Her b            &        K0V & $ 18.15\pm  0.19$ &     26.92 & 27.75$^{+0.44}_{-0.38}$ & 33.43 &    7.45  &  4.64 &   2.77 & $ -1.41$ & $ -0.53 $ & 17.52 &    18.28 &      4.4e+05 &      3.3e+06 \\
16 Cyg B b          &      G2.5V & $ 21.62\pm  0.23$ &  $<$26.71 &                $<$28.04 & 33.78 &   10.31  &  1.68 &   1.68 & $<-1.19$ & $< 0.16 $ & 18.04 &    18.74 &    (7.2e+05) &    (1.6e+07) \\
2M1207 b            &         M8 & $ 52.41\pm  1.10$ &  $<$26.24 &                $<$27.16 & 30.07 &  $ <15$  &  4.00 &  46.00 & $<-4.53$ & $<-3.56 $ & 13.71 &    15.78 &    (3.3e+02) &    (3.1e+03) \\
30 Ari B b          &        F6V & $ 39.43\pm  1.71$ &     29.53 & 29.36$^{+0.07}_{-0.03}$ & 33.83 &    0.16  &  9.88 &   1.00 & $  2.09$ & $  2.31 $ & 18.35 &    18.93 &      1.4e+09 &      2.3e+09 \\
47 UMa b            &        G0V & $ 14.08\pm  0.13$ &     25.45 & 26.56$^{+0.34}_{-0.22}$ & 33.74 &  $ <15$  &  2.53 &   2.10 & $ -2.64$ & $ -1.50 $ & 17.84 &    18.55 &      2.6e+04 &      3.6e+05 \\
47 UMa c            &            &                   &           &                         &       &          &  0.54 &   3.60 & $ -3.11$ & $ -1.97$ & 17.37  &    18.08 &      8.8e+03 &      1.2e+05 \\
47 UMa d            &            &                   &           &                         &       &          &  1.64 &  11.60 & $ -4.12$ & $ -2.98$ & 16.36  &    17.06 &      8.4e+02 &      1.2e+04 \\
51 Peg b            &       G2IV & $ 15.36\pm  0.18$ &  $<$26.52 &                $<$27.86 & 33.65 &   13.50: &  0.47 &   0.05 & $< 1.64$ & $< 3.00$ & 21.04: &   21.76: &    (4.9e+08) &    (1.1e+10) \\
55 Cnc b            &        G8V & $ 12.53\pm  0.13$ &     26.65 & 27.49$^{+0.46}_{-0.42}$ & 33.39 &   11.19  &  0.82 &   0.12 & $  1.08$ & $  1.98 $ & 20.28 &    21.05 &      1.4e+08 &      1.1e+09 \\
55 Cnc c            &            &                   &           &                         &       &          &  0.17 &   0.24 & $  0.44$ & $  1.34$ & 19.64  &    20.41 &      3.1e+07 &      2.4e+08 \\
55 Cnc d            &            &                   &           &                         &       &          &  3.84 &   5.77 & $ -2.32$ & $ -1.42$ & 16.88  &    17.65 &      5.4e+04 &      4.2e+05 \\
55 Cnc e            &            &                   &           &                         &       &          &  0.02 &   0.04 & $  2.04$ & $  2.94$ & 21.24  &    22.01 &      1.2e+09 &      9.7e+09 \\
55 Cnc f            &            &                   &           &                         &       &          &  0.14 &   0.79 & $ -0.59$ & $  0.31$ & 18.61  &    19.38 &      2.9e+06 &      2.3e+07 \\
$\beta$ Pic b       &        A6V & $ 19.28\pm  0.19$ &     25.63 & 27.01$^{+0.44}_{-0.39}$ & 34.49 &   \dots  &  8.00 &  12.00 & $ -3.97$ & $ -2.58 $ & \dots &    \dots &      1.2e+03 &      2.9e+04 \\
$\epsilon$ Eridani b&        K2V & $  3.22\pm  0.01$ &     28.20 & 28.44$^{+0.00}_{-0.00}$ & 33.10 &    1.12  &  1.55 &   3.39 & $ -0.31$ & $  0.13 $ & 17.19 &    18.00 &      5.5e+06 &      1.5e+07 \\
GJ 86 b             &        K1V & $ 10.91\pm  0.07$ &  $<$27.42 &                $<$29.15 & 33.16 &    3.59  &  4.01 &   0.11 & $< 1.88$ & $< 3.63 $ & 20.23 &    21.04 &    (8.6e+08) &    (4.8e+10) \\
GJ 317 b            &       M3.5 & $  9.01\pm  0.97$ &     26.12 & 27.43$^{+0.47}_{-0.45}$ & 30.18 &  $ <15$  &  1.20 &   0.95 & $ -1.28$ & $  0.05 $ & 17.15 &    19.15 &      5.9e+05 &      1.3e+07 \\
GJ 436 b            &       M2.5 & $ 10.23\pm  0.24$ &     25.96 & 26.87$^{+0.45}_{-0.40}$ & 31.67 &  $ <15$  &  0.07 &   0.03 & $  1.59$ & $  2.55 $ & 20.88 &    22.20 &      4.4e+08 &      4.0e+09 \\
GJ 674 b            &       M2.5 & $  4.54\pm  0.03$ &     27.72 & 28.01$^{+0.28}_{-0.17}$ & 31.54 &    2.27  &  0.04 &   0.04 & $  3.09$ & $  3.56 $ & 20.42 &    21.86 &      1.4e+10 &      4.1e+10 \\
GJ 876 b            &        M4V & $  4.70\pm  0.05$ &     26.16 & 26.65$^{+0.41}_{-0.31}$ & 31.37 &  $ <15$  &  2.28 &   0.21 & $  0.07$ & $  0.69 $ & 19.04 &    20.48 &      1.3e+07 &      5.5e+07 \\
GJ 876 c            &            &                   &           &                         &       &          &  0.71 &   0.13 & $  0.48$ & $  1.10$ & 19.45  &    20.89 &      3.4e+07 &      1.4e+08 \\
GJ 876 d            &            &                   &           &                         &       &          &  0.02 &   0.02 & $  2.07$ & $  2.69$ & 21.04  &    22.48 &      1.3e+09 &      5.5e+09 \\
GJ 876 e            &            &                   &           &                         &       &          &  0.05 &   0.33 & $ -0.34$ & $  0.28$ & 18.63  &    20.07 &      5.2e+06 &      2.1e+07 \\
GQ Lup b            &       K7V  & $140:$            &     29.45 & 31.05$^{+0.47}_{-0.45}$ & 33.36 &    0.17  & 21.50 & 103.00 & $ -2.02$ & $ -0.41 $ & 14.16 &    14.87 &      1.1e+05 &      4.3e+06 \\
HD 4308 b           &        G5V & $ 21.85\pm  0.27$ &  $<$25.94 &                $<$27.28 & 33.57 &  $ <15$  &  0.04 &   0.12 & $< 0.35$ & $< 1.71 $ & 20.31 &    21.04 &    (2.5e+07) &    (5.7e+08) \\
HD 20367 b          &        G0V & $ 27.13\pm  0.79$ &     29.29 & 29.43$^{+0.33}_{-0.21}$ & 33.78 &    0.22  &  1.07 &   1.25 & $  1.65$ & $  2.02 $ & 18.17 &    18.77 &      5.0e+08 &      1.2e+09 \\
HD 27442 b          &      K2IV  & $ 18.23\pm  0.17$ &  $<$26.14 &                $<$27.47 & 34.40 &  $ <15$  &  1.35 &   1.16 & $<-1.44$ & $<-0.09$ & 18.58: &   19.12: &    (4.1e+05) &    (9.2e+06) \\
HD 46375 b          &       K1IV & $ 33.41\pm  1.19$ &     27.16 & 28.49$^{+0.47}_{-0.45}$ & 33.48 &    5.22: &  0.25 &   0.04 & $  2.49$ & $  3.84$ & 21.19: &   21.93: &      3.5e+09 &      7.7e+10 \\
HD 49674 b          &        G5V & $ 40.73\pm  1.89$ &     27.41 & 29.36$^{+0.19}_{-0.08}$ & 33.51 &    3.62  &  0.12 &   0.06 & $  2.43$ & $  4.39 $ & 20.89 &    21.61 &      3.1e+09 &      2.8e+11 \\
HD 50554 b          &         F8 & $ 31.03\pm  0.97$ &  $<$26.32 &                $<$27.66 & 33.74 &  $ <15$  &  5.16 &   2.41 & $<-1.89$ & $<-0.53 $ & 17.72 &    18.43 &    (1.4e+05) &    (3.3e+06) \\
HD 52265 b          &        G0V & $ 28.07\pm  0.66$ &     26.89 & 27.72$^{+0.38}_{-0.28}$ & 33.86 &    7.80  &  1.05 &   0.50 & $  0.05$ & $  0.93 $ & 19.11 &    19.78 &      1.3e+07 &      9.7e+07 \\
HD 70642 b          &     G5IV-V & $ 28.76\pm  0.50$ &     26.38 & 27.71$^{+0.47}_{-0.44}$ & 33.58 &  $ <15$  &  2.00 &   3.30 & $ -2.10$ & $ -0.76$ & 17.41: &   18.15: &      8.9e+04 &      2.0e+06 \\
HD 75289 b          &        G0V & $ 28.94\pm  0.47$ &  $<$25.90 &                $<$27.24 & 33.87 &  $ <15$  &  0.42 &   0.05 & $< 1.13$ & $< 2.49 $ & 21.19 &    21.87 &    (1.5e+08) &    (3.4e+09) \\
HD 93083 b          &        K3V & $ 28.90\pm  0.84$ &     26.90 & 28.36$^{+0.47}_{-0.46}$ & 33.19 &    7.77  &  0.37 &   0.48 & $  0.09$ & $  1.56 $ & 18.98 &    19.79 &      1.4e+07 &      4.1e+08 \\
HD 95089 b          &       K0IV & $139.08\pm  0.00$ &  $<$27.09 &                $<$28.43 & 34.71 &    5.85: &  1.20 &   1.51 & $<-0.72$ & $< 0.64$ & 18.46: &   18.90: &    (2.1e+06) &    (4.9e+07) \\
HD 99492 b          &        K2V & $ 17.99\pm  1.07$ &     26.55 & 27.85$^{+0.47}_{-0.45}$ & 33.11 &   13.01  &  0.11 &   0.12 & $  0.92$ & $  2.24 $ & 20.14 &    20.98 &      9.3e+07 &      2.0e+09 \\
HD 99492 c          &            &                   &           &                         &       &          &  0.36 &   5.40 & $ -2.36$ & $ -1.04$ & 16.85  &    17.70 &      4.9e+04 &      1.0e+06 \\
HD 101930 b         &        K1V & $ 30.50\pm  0.89$ &     26.05 & 27.00$^{+0.39}_{-0.28}$ & 33.27 &  $ <15$  &  0.30 &   0.30 & $ -0.36$ & $  0.64 $ & 19.41 &    20.21 &      4.9e+06 &      4.9e+07 \\
HD 102195 b         &        K0V & $ 28.98\pm  0.97$ &     28.43 & 29.76$^{+0.41}_{-0.32}$ & 33.26 &    0.80  &  0.45 &   0.05 & $  3.60$ & $  4.95 $ & 20.91 &    21.67 &      4.5e+10 &      1.0e+12 \\
HD 108147 b         &     F8/G0V & $ 38.57\pm  1.03$ &     27.39 & 28.67$^{+0.47}_{-0.45}$ & 33.85 &    3.74  &  0.26 &   0.10 & $  1.92$ & $  3.23 $ & 20.48 &    21.14 &      9.4e+08 &      1.9e+10 \\
HD 111232 b         &        G8V & $ 28.88\pm  0.67$ &  $<$26.35 &                $<$27.69 & 33.40 &  $ <15$  &  6.80 &   1.97 & $<-1.69$ & $<-0.33 $ & 17.81 &    18.59 &    (2.3e+05) &    (5.3e+06) \\
HD 114386 b         &        K3V & $ 28.04\pm  1.04$ &     26.53 & 27.40$^{+0.42}_{-0.34}$ & 33.02 &   13.41  &  1.24 &   1.65 & $ -1.36$ & $ -0.43 $ & 17.85 &    18.72 &      5.0e+05 &      4.1e+06 \\
HD 114762 b         &        F9V & $ 40.57\pm  2.37$ &  $<$26.51 &                $<$27.83 & 33.76 &   13.73  & 11.68 &   0.36 & $<-0.06$ & $< 1.28 $ & 19.37 &    20.07 &    (9.9e+06) &    (2.1e+08) \\
HD 114783 b         &        K0V & $ 20.43\pm  0.44$ &  $<$26.54 &                $<$27.83 & 33.19 &   13.17  &  1.00 &   1.20 & $<-1.07$ & $< 0.24 $ & 18.18 &    19.01 &    (9.6e+05) &    (2.0e+07) \\
HD 130322 b         &        K0V & $ 29.76\pm  1.34$ &     27.25 & 28.57$^{+0.47}_{-0.44}$ & 33.28 &    4.60  &  1.02 &   0.09 & $  1.91$ & $  3.25 $ & 20.47 &    21.25 &      9.2e+08 &      2.0e+10 \\
HD 154345 b         &        G8V & $ 18.06\pm  0.18$ &     27.12 & 27.95$^{+0.37}_{-0.26}$ & 33.36 &    5.60  &  0.95 &   4.19 & $ -1.58$ & $ -0.69 $ & 17.14 &    17.90 &      3.0e+05 &      2.3e+06 \\
HD 164922 b         &        K0V & $ 21.93\pm  0.34$ &  $<$25.74 &                $<$27.08 & 33.44 &  $ <15$  &  0.36 &   2.11 & $<-2.35$ & $<-1.00 $ & 17.77 &    18.53 &    (5.0e+04) &    (1.1e+06) \\
HD 179949 b         &        F8V & $ 27.05\pm  0.59$ &     28.38 & 29.52$^{+0.39}_{-0.29}$ & 33.84 &    0.85  &  0.95 &   0.05 & $  3.63$ & $  4.80 $ & 21.15 &    21.78 &      4.8e+10 &      7.0e+11 \\
HD 187123 b         &         G5 & $ 47.92\pm  1.63$ &  $<$27.24 &                $<$28.62 & 33.78 &    4.67  &  0.52 &   0.04 & $< 2.53$ & $< 3.93 $ & 21.22 &    21.90 &    (3.8e+09) &    (9.6e+10) \\
HD 187123 c         &            &                   &           &                         &       &          &  1.99 &   4.89 & $<-1.59$ & $<-0.19$ & 17.11  &    17.78 &    (2.9e+05) &    (7.3e+06) \\
HD 189733 b         &      K1-K2 & $ 19.25\pm  0.32$ &     28.18 & 28.48$^{+0.18}_{-0.08}$ & 33.10 &    1.16  &  1.15 &   0.03 & $  3.73$ & $  4.21 $ & 21.26 &    22.07 &      6.1e+10 &      1.8e+11 \\
HD 190360 b         &       G6IV & $ 15.89\pm  0.16$ &  $<$26.35 &                $<$27.69 & 33.65 &  $ <15$  &  1.50 &   3.92 & $<-2.29$ & $<-0.93$ & 17.28: &   18.00: &    (5.8e+04) &    (1.3e+06) \\
HD 190360 c         &            &                   &           &                         &       &          &  0.06 &   0.13 & $< 0.69$ & $< 2.05$ & 20.25: &   20.98: &    (5.5e+07) &    (1.3e+09) \\
HD 195019 b         &     G3IV-V & $ 37.36\pm  1.24$ &  $<$26.23 &                $<$27.57 & 33.89 &  $ <15$  &  3.70 &   0.14 & $< 0.50$ & $< 1.86$ & 20.24: &   20.92: &    (3.5e+07) &    (8.1e+08) \\
HD 209458 b         &        G0V & $ 47.08\pm  2.22$ &  $<$26.40 &                $<$27.74 & 33.78 &  $ <15$  &  0.71 &   0.05 & $< 1.60$ & $< 2.96 $ & 21.14 &    21.84 &    (4.5e+08) &    (1.0e+10) \\
HD 216435 b         &        G0V & $ 33.29\pm  0.81$ &     27.74 & 28.99$^{+0.47}_{-0.44}$ & 34.14 &    2.22  &  1.26 &   2.56 & $ -0.53$ & $  0.75 $ & 17.78 &    18.35 &      3.3e+06 &      6.3e+07 \\
HD 216437 b         &     G4IV-V & $ 26.52\pm  0.41$ &     26.62 & 27.47$^{+0.42}_{-0.35}$ & 33.92 &   11.68: &  1.82 &   2.32 & $ -1.56$ & $ -0.65$ & 17.80: &   18.47: &      3.1e+05 &      2.5e+06 \\
HD 217107 b         &       G8IV & $ 19.72\pm  0.29$ &  $<$25.50 &                $<$26.84 & 33.64 &  $ <15$  &  1.33 &   0.07 & $< 0.32$ & $< 1.68$ & 20.74: &   21.46: &    (2.4e+07) &    (5.4e+08) \\
HD 217107 c         &            &                   &           &                         &       &          &  2.49 &   5.27 & $<-3.40$ & $<-2.03$ & 17.02: &   17.75: &    (4.5e+03) &    (1.0e+05) \\
HD 218566 b         &        K3V & $ 29.94\pm  1.07$ &     27.04 & 28.35$^{+0.47}_{-0.44}$ & 33.13 &    6.28  &  0.21 &   0.69 & $ -0.08$ & $  1.25 $ & 18.64 &    19.47 &      9.3e+06 &      2.0e+08 \\
HD 330075 b         &         G5 & $ 50.20\pm  3.75$ &     26.51 & 27.36$^{+0.46}_{-0.42}$ & 33.26 &   13.80  &  0.62 &   0.04 & $  1.88$ & $  2.79 $ & 21.18 &    21.99 &      8.5e+08 &      6.9e+09 \\
HR 8799 b           &        A5V & $ 39.94\pm  1.36$ &     28.02 & 29.37$^{+0.47}_{-0.45}$ & 34.26 &   \dots  &  7.00 &  68.00 & $ -3.09$ & $ -1.72 $ & \dots &    \dots &      9.0e+03 &      2.1e+05 \\
HR 8799 c           &            &                   &           &                         &       &          & 10.00 &  38.00 & $ -2.59$ & $ -1.22 $ & \dots &    \dots &      2.9e+04 &      6.8e+05 \\
HR 8799 d           &            &                   &           &                         &       &          & 10.00 &  24.00 & $ -2.19$ & $ -0.82 $ & \dots &    \dots &      7.3e+04 &      1.7e+06 \\
HR 8799 e           &            &                   &           &                         &       &          &  9.00 &  14.50 & $ -1.75$ & $ -0.38 $ & \dots &    \dots &      2.0e+05 &      4.7e+06 \\
$\mu$ Ara b         &     G3IV-V & $ 15.28\pm  0.19$ &  $<$25.99 &                $<$27.32 & 33.83 &  $ <15$  &  1.68 &   1.50 & $<-1.81$ & $<-0.46$ & 18.15: &   18.84: &    (1.7e+05) &    (3.9e+06) \\
$\mu$ Ara c         &            &                   &           &                         &       &          &  0.03 &   0.09 & $< 0.62$ & $< 1.97$ & 20.59: &   21.28: &    (4.7e+07) &    (1.1e+09) \\
$\mu$ Ara d         &            &                   &           &                         &       &          &  0.52 &   0.92 & $<-1.39$ & $<-0.04$ & 18.58: &   19.27: &    (4.6e+05) &    (1.0e+07) \\
$\mu$ Ara e         &            &                   &           &                         &       &          &  1.81 &   5.24 & $<-2.90$ & $<-1.55$ & 17.07: &   17.76: &    (1.4e+04) &    (3.2e+05) \\
NGC 2423 3 b        &            & $766:$            &  $<$29.27 &                $<$34.91 & 35.54 &    0.23  & 10.60 &   2.10 & $< 1.18$ & $< 6.82 $ & 18.45 &    18.64 &    (1.7e+08) &    (7.4e+13) \\
Pollux b            &     K0III  & $ 10.34\pm  0.09$ &     27.13 & 28.38$^{+0.21}_{-0.10}$ & 35.19 &   \dots  &  2.90 &   1.69 & $ -0.77$ & $  0.50 $ & \dots &    \dots &      1.9e+06 &      3.6e+07 \\
$\tau$ Boo b        &        F7V & $ 15.60\pm  0.17$ &     28.95 & 29.40$^{+0.39}_{-0.28}$ & 34.06 &    0.37  &  3.90 &   0.05 & $  4.18$ & $  4.76 $ & 21.19 &    21.72 &      1.7e+11 &      6.4e+11 \\
$\upsilon$ And b    &        F8V & $ 13.47\pm  0.13$ &     27.56 & 28.73$^{+0.46}_{-0.41}$ & 34.11 &    2.88  &  0.69 &   0.06 & $  2.57$ & $  3.77 $ & 21.04 &    21.63 &      4.2e+09 &      6.6e+10 \\
$\upsilon$ And c    &            &                   &           &                         &       &          & 14.57 &   0.86 & $  0.25$ & $  1.44$ & 18.72  &    19.30 &      2.0e+07 &      3.1e+08 \\
$\upsilon$ And d    &            &                   &           &                         &       &          & 10.19 &   2.55 & $ -0.70$ & $  0.50$ & 17.77  &    18.36 &      2.3e+06 &      3.6e+07 \\
$\upsilon$ And e    &            &                   &           &                         &       &          &  1.06 &   5.25 & $ -1.32$ & $ -0.13$ & 17.15  &    17.73 &      5.3e+05 &      8.4e+06 \\
%
%---------------------------------------------
\hline
\end{tabular}
\end{scriptsize}
\end{center}
\vspace{-3mm}
 \tablefoot{
 \tablefoottext{a}{Planet data from The Extrasolar Planets Encyclopedia (http://exoplanet.eu).} 
 \tablefoottext{c}{1~M$_{\rm J}$\,Gyr$^{-1}$=$6.02\times 10^{13}$~g\,s$^{-1}$}
}
\end{table*}
%}
%---------------------------------------------

%% file: tablecoolemds.tex
%------------------------------------------------- begin table
%  Emission Measure Distribution in transition region
\begin{landscape}
\begin{table}
\caption{Emission measure distribution in the transition region}\label{tab:coolemd}
\tabcolsep 2.pt
\centering
\begin{scriptsize}
\begin{tabular}{lcccccccccccccccccc}
\hline \hline
Star & \multicolumn{18}{c}{EM (cm$^{-3}$)\tablefootmark{a}} \\ 
     & $\log T$\,(K)=4.0 & 4.1 & 4.2 & 4.3 & 4.4 & 4.5 & 4.6 & 4.7 & 4.8 & 4.9 & 5.0 & 5.1 & 5.2 & 5.3 & 5.4 & 5.5 & 5.6 & 5.7 \\
\hline
%----------------
14 Her        & $49.96^{+0.62}_{-0.61}$ & $49.80^{+0.60}_{-0.60}$ & $49.63^{+0.59}_{-0.58}$ & $49.46^{+0.57}_{-0.57}$ & $49.30^{+0.55}_{-0.56}$ & $49.13^{+0.54}_{-0.54}$ & $48.97^{+0.52}_{-0.53}$ & $48.80^{+0.51}_{-0.51}$ & $48.63^{+0.50}_{-0.49}$ & $48.47^{+0.48}_{-0.48}$ & $48.30^{+0.47}_{-0.46}$ & $48.14^{+0.45}_{-0.45}$ & $47.97^{+0.44}_{-0.43}$ & $47.80^{+0.42}_{-0.42}$ & $47.64^{+0.40}_{-0.41}$ & $47.47^{+0.39}_{-0.39}$ & $47.31^{+0.37}_{-0.38}$ & $47.14^{+0.36}_{-0.36}$ \\
 16 Cyg B     & $<50.40$ & $<50.22$ & $<50.04$ & $<49.85$ & $<49.67$ & $<49.49$ & $<49.31$ & $<49.13$ & $<48.95$ & $<48.77$ & $<48.59$ & $<48.41$ & $<48.23$ & $<48.04$ & $<47.86$ & $<47.68$ & $<47.50$ & $<47.32$ \\
2M 1207 A     & $<49.52$ & $<49.34$ & $<49.16$ & $<48.97$ & $<48.79$ & $<48.61$ & $<48.43$ & $<48.25$ & $<48.07$ & $<47.89$ & $<47.71$ & $<47.53$ & $<47.35$ & $<47.16$ & $<46.98$ & $<46.80$ & $<46.62$ & $<46.44$ \\
30 Ari B      & $50.66^{+0.69}_{-0.69}$ & $50.51^{+0.66}_{-0.67}$ & $50.36^{+0.64}_{-0.64}$ & $50.20^{+0.62}_{-0.61}$ & $50.05^{+0.59}_{-0.59}$ & $49.90^{+0.56}_{-0.57}$ & $49.74^{+0.54}_{-0.53}$ & $49.59^{+0.51}_{-0.51}$ & $49.44^{+0.48}_{-0.49}$ & $49.28^{+0.46}_{-0.45}$ & $49.13^{+0.43}_{-0.43}$ & $48.98^{+0.40}_{-0.41}$ & $48.83^{+0.38}_{-0.39}$ & $48.67^{+0.36}_{-0.35}$ & $48.52^{+0.33}_{-0.33}$ & $48.37^{+0.30}_{-0.31}$ & $48.21^{+0.28}_{-0.27}$ & $48.06^{+0.25}_{-0.25}$ \\
47 UMa        & $48.61^{+0.62}_{-0.61}$ & $48.45^{+0.60}_{-0.60}$ & $48.28^{+0.59}_{-0.59}$ & $48.11^{+0.57}_{-0.57}$ & $47.95^{+0.55}_{-0.56}$ & $47.78^{+0.54}_{-0.54}$ & $47.62^{+0.52}_{-0.53}$ & $47.45^{+0.51}_{-0.51}$ & $47.28^{+0.50}_{-0.49}$ & $47.12^{+0.48}_{-0.48}$ & $46.95^{+0.47}_{-0.46}$ & $46.79^{+0.45}_{-0.45}$ & $46.62^{+0.44}_{-0.43}$ & $46.45^{+0.42}_{-0.42}$ & $46.29^{+0.40}_{-0.41}$ & $46.12^{+0.39}_{-0.39}$ & $45.96^{+0.37}_{-0.38}$ & $45.79^{+0.36}_{-0.36}$ \\
 51 Peg        & $<50.22$ & $<50.04$ & $<49.86$ & $<49.67$ & $<49.49$ & $<49.31$ & $<49.13$ & $<48.95$ & $<48.77$ & $<48.59$ & $<48.41$ & $<48.23$ & $<48.05$ & $<47.86$ & $<47.68$ & $<47.50$ & $<47.32$ & $<47.14$ \\
55 Cnc       & $49.73^{+0.62}_{-0.61}$ & $49.57^{+0.60}_{-0.60}$ & $49.40^{+0.59}_{-0.59}$ & $49.23^{+0.57}_{-0.57}$ & $49.07^{+0.55}_{-0.56}$ & $48.90^{+0.54}_{-0.54}$ & $48.74^{+0.52}_{-0.53}$ & $48.57^{+0.51}_{-0.51}$ & $48.40^{+0.50}_{-0.49}$ & $48.24^{+0.48}_{-0.48}$ & $48.07^{+0.47}_{-0.46}$ & $47.91^{+0.45}_{-0.45}$ & $47.74^{+0.43}_{-0.43}$ & $47.57^{+0.42}_{-0.42}$ & $47.41^{+0.40}_{-0.41}$ & $47.24^{+0.39}_{-0.39}$ & $47.08^{+0.37}_{-0.38}$ & $46.91^{+0.36}_{-0.36}$ \\
$\beta$ Pic    & $49.23^{+0.62}_{-0.61}$ & $49.07^{+0.60}_{-0.60}$ & $48.90^{+0.59}_{-0.59}$ & $48.73^{+0.57}_{-0.57}$ & $48.57^{+0.55}_{-0.56}$ & $48.40^{+0.54}_{-0.54}$ & $48.24^{+0.52}_{-0.53}$ & $48.07^{+0.51}_{-0.51}$ & $47.90^{+0.50}_{-0.49}$ & $47.74^{+0.48}_{-0.48}$ & $47.57^{+0.47}_{-0.46}$ & $47.41^{+0.45}_{-0.45}$ & $47.24^{+0.43}_{-0.43}$ & $47.07^{+0.42}_{-0.42}$ & $46.91^{+0.40}_{-0.41}$ & $46.74^{+0.39}_{-0.39}$ & $46.58^{+0.37}_{-0.38}$ & $46.41^{+0.36}_{-0.36}$ \\
 GJ 317       & $49.69^{+0.62}_{-0.61}$ & $49.53^{+0.60}_{-0.60}$ & $49.36^{+0.59}_{-0.58}$ & $49.19^{+0.57}_{-0.57}$ & $49.03^{+0.55}_{-0.56}$ & $48.86^{+0.54}_{-0.54}$ & $48.70^{+0.52}_{-0.53}$ & $48.53^{+0.51}_{-0.51}$ & $48.36^{+0.50}_{-0.49}$ & $48.20^{+0.48}_{-0.48}$ & $48.03^{+0.47}_{-0.46}$ & $47.87^{+0.45}_{-0.45}$ & $47.70^{+0.44}_{-0.43}$ & $47.53^{+0.42}_{-0.42}$ & $47.37^{+0.40}_{-0.41}$ & $47.20^{+0.39}_{-0.39}$ & $47.04^{+0.37}_{-0.38}$ & $46.87^{+0.36}_{-0.36}$ \\
GJ 436        & $49.10^{+0.62}_{-0.61}$ & $48.94^{+0.60}_{-0.60}$ & $48.77^{+0.59}_{-0.58}$ & $48.60^{+0.57}_{-0.57}$ & $48.44^{+0.55}_{-0.56}$ & $48.27^{+0.54}_{-0.54}$ & $48.11^{+0.52}_{-0.53}$ & $47.94^{+0.51}_{-0.51}$ & $47.77^{+0.50}_{-0.49}$ & $47.61^{+0.48}_{-0.48}$ & $47.44^{+0.47}_{-0.46}$ & $47.28^{+0.45}_{-0.45}$ & $47.11^{+0.44}_{-0.43}$ & $46.94^{+0.42}_{-0.42}$ & $46.78^{+0.40}_{-0.41}$ & $46.61^{+0.39}_{-0.39}$ & $46.45^{+0.37}_{-0.38}$ & $46.28^{+0.36}_{-0.36}$ \\
 GJ 674       & $50.05^{+0.69}_{-0.69}$ & $49.90^{+0.66}_{-0.67}$ & $49.75^{+0.64}_{-0.65}$ & $49.59^{+0.62}_{-0.61}$ & $49.44^{+0.59}_{-0.59}$ & $49.29^{+0.56}_{-0.57}$ & $49.13^{+0.54}_{-0.53}$ & $48.98^{+0.51}_{-0.51}$ & $48.83^{+0.48}_{-0.49}$ & $48.67^{+0.46}_{-0.45}$ & $48.52^{+0.43}_{-0.43}$ & $48.37^{+0.40}_{-0.41}$ & $48.22^{+0.38}_{-0.39}$ & $48.06^{+0.36}_{-0.35}$ & $47.91^{+0.33}_{-0.33}$ & $47.76^{+0.30}_{-0.31}$ & $47.60^{+0.28}_{-0.27}$ & $47.45^{+0.25}_{-0.25}$ \\
 GJ 86        & $<51.56$ & $<51.38$ & $<51.20$ & $<51.01$ & $<50.83$ & $<50.65$ & $<50.47$ & $<50.29$ & $<50.11$ & $<49.93$ & $<49.75$ & $<49.57$ & $<49.39$ & $<49.20$ & $<49.02$ & $<48.84$ & $<48.66$ & $<48.48$ \\
 GJ 876       & $48.81^{+0.62}_{-0.61}$ & $48.65^{+0.60}_{-0.60}$ & $48.48^{+0.59}_{-0.58}$ & $48.31^{+0.57}_{-0.57}$ & $48.15^{+0.55}_{-0.56}$ & $47.98^{+0.54}_{-0.54}$ & $47.82^{+0.52}_{-0.53}$ & $47.65^{+0.51}_{-0.51}$ & $47.48^{+0.50}_{-0.49}$ & $47.32^{+0.48}_{-0.48}$ & $47.15^{+0.47}_{-0.46}$ & $46.99^{+0.45}_{-0.45}$ & $46.82^{+0.44}_{-0.43}$ & $46.65^{+0.42}_{-0.42}$ & $46.49^{+0.40}_{-0.41}$ & $46.32^{+0.39}_{-0.39}$ & $46.16^{+0.37}_{-0.38}$ & $45.99^{+0.36}_{-0.36}$ \\
GQ Lup        & $53.31^{+0.62}_{-0.61}$ & $53.15^{+0.60}_{-0.60}$ & $52.98^{+0.59}_{-0.58}$ & $52.81^{+0.57}_{-0.57}$ & $52.65^{+0.55}_{-0.56}$ & $52.48^{+0.54}_{-0.54}$ & $52.32^{+0.52}_{-0.53}$ & $52.15^{+0.51}_{-0.51}$ & $51.98^{+0.50}_{-0.49}$ & $51.82^{+0.48}_{-0.48}$ & $51.65^{+0.47}_{-0.46}$ & $51.49^{+0.45}_{-0.45}$ & $51.32^{+0.44}_{-0.43}$ & $51.15^{+0.42}_{-0.42}$ & $50.99^{+0.40}_{-0.41}$ & $50.82^{+0.39}_{-0.39}$ & $50.66^{+0.37}_{-0.38}$ & $50.49^{+0.36}_{-0.36}$ \\
 HD 4308      & $<49.64$ & $<49.46$ & $<49.28$ & $<49.09$ & $<48.91$ & $<48.73$ & $<48.55$ & $<48.37$ & $<48.19$ & $<48.01$ & $<47.83$ & $<47.65$ & $<47.47$ & $<47.28$ & $<47.10$ & $<46.92$ & $<46.74$ & $<46.56$ \\
HD 20367      & $51.59^{+0.62}_{-0.61}$ & $51.43^{+0.60}_{-0.60}$ & $51.26^{+0.59}_{-0.59}$ & $51.09^{+0.57}_{-0.57}$ & $50.93^{+0.55}_{-0.56}$ & $50.76^{+0.54}_{-0.54}$ & $50.60^{+0.52}_{-0.53}$ & $50.43^{+0.51}_{-0.51}$ & $50.26^{+0.50}_{-0.49}$ & $50.10^{+0.48}_{-0.48}$ & $49.93^{+0.47}_{-0.46}$ & $49.77^{+0.45}_{-0.45}$ & $49.60^{+0.43}_{-0.43}$ & $49.43^{+0.42}_{-0.42}$ & $49.27^{+0.40}_{-0.41}$ & $49.10^{+0.39}_{-0.39}$ & $48.94^{+0.37}_{-0.38}$ & $48.77^{+0.36}_{-0.36}$ \\
 HD 27442     & $<49.83$ & $<49.65$ & $<49.47$ & $<49.28$ & $<49.10$ & $<48.92$ & $<48.74$ & $<48.56$ & $<48.38$ & $<48.20$ & $<48.02$ & $<47.84$ & $<47.66$ & $<47.47$ & $<47.29$ & $<47.11$ & $<46.93$ & $<46.75$ \\
HD 46375      & $50.75^{+0.62}_{-0.61}$ & $50.59^{+0.60}_{-0.60}$ & $50.42^{+0.59}_{-0.59}$ & $50.25^{+0.57}_{-0.57}$ & $50.09^{+0.55}_{-0.56}$ & $49.92^{+0.54}_{-0.54}$ & $49.76^{+0.52}_{-0.53}$ & $49.59^{+0.51}_{-0.51}$ & $49.42^{+0.50}_{-0.49}$ & $49.26^{+0.48}_{-0.48}$ & $49.09^{+0.47}_{-0.46}$ & $48.93^{+0.45}_{-0.45}$ & $48.76^{+0.43}_{-0.43}$ & $48.59^{+0.42}_{-0.42}$ & $48.43^{+0.40}_{-0.41}$ & $48.26^{+0.39}_{-0.39}$ & $48.10^{+0.37}_{-0.38}$ & $47.93^{+0.36}_{-0.36}$ \\
HD 49674      & $51.21^{+0.62}_{-0.61}$ & $51.05^{+0.60}_{-0.60}$ & $50.88^{+0.59}_{-0.58}$ & $50.71^{+0.57}_{-0.57}$ & $50.55^{+0.55}_{-0.56}$ & $50.38^{+0.54}_{-0.54}$ & $50.22^{+0.52}_{-0.53}$ & $50.05^{+0.51}_{-0.51}$ & $49.88^{+0.50}_{-0.49}$ & $49.72^{+0.48}_{-0.48}$ & $49.55^{+0.47}_{-0.46}$ & $49.39^{+0.45}_{-0.45}$ & $49.22^{+0.44}_{-0.43}$ & $49.05^{+0.42}_{-0.42}$ & $48.89^{+0.40}_{-0.41}$ & $48.72^{+0.39}_{-0.39}$ & $48.56^{+0.37}_{-0.38}$ & $48.39^{+0.36}_{-0.36}$ \\
HD 50554      & $<50.02$ & $<49.84$ & $<49.66$ & $<49.47$ & $<49.29$ & $<49.11$ & $<48.93$ & $<48.75$ & $<48.57$ & $<48.39$ & $<48.21$ & $<48.03$ & $<47.85$ & $<47.66$ & $<47.48$ & $<47.30$ & $<47.12$ & $<46.94$ \\
 HD 52265     & $49.84^{+0.62}_{-0.61}$ & $49.68^{+0.60}_{-0.60}$ & $49.51^{+0.59}_{-0.59}$ & $49.34^{+0.57}_{-0.57}$ & $49.18^{+0.55}_{-0.56}$ & $49.01^{+0.54}_{-0.54}$ & $48.85^{+0.52}_{-0.53}$ & $48.68^{+0.51}_{-0.51}$ & $48.51^{+0.50}_{-0.49}$ & $48.35^{+0.48}_{-0.48}$ & $48.18^{+0.47}_{-0.46}$ & $48.02^{+0.45}_{-0.45}$ & $47.85^{+0.43}_{-0.43}$ & $47.68^{+0.42}_{-0.42}$ & $47.52^{+0.40}_{-0.41}$ & $47.35^{+0.39}_{-0.39}$ & $47.19^{+0.37}_{-0.38}$ & $47.02^{+0.36}_{-0.36}$ \\
HD 70642      & $49.96^{+0.62}_{-0.61}$ & $49.80^{+0.60}_{-0.60}$ & $49.63^{+0.59}_{-0.58}$ & $49.46^{+0.57}_{-0.57}$ & $49.30^{+0.55}_{-0.56}$ & $49.13^{+0.54}_{-0.54}$ & $48.97^{+0.52}_{-0.53}$ & $48.80^{+0.51}_{-0.51}$ & $48.63^{+0.50}_{-0.49}$ & $48.47^{+0.48}_{-0.48}$ & $48.30^{+0.47}_{-0.46}$ & $48.14^{+0.45}_{-0.45}$ & $47.97^{+0.44}_{-0.43}$ & $47.80^{+0.42}_{-0.42}$ & $47.64^{+0.40}_{-0.41}$ & $47.47^{+0.39}_{-0.39}$ & $47.31^{+0.37}_{-0.38}$ & $47.14^{+0.36}_{-0.36}$ \\
HD 75289      & $<49.60$ & $<49.42$ & $<49.24$ & $<49.05$ & $<48.87$ & $<48.69$ & $<48.51$ & $<48.33$ & $<48.15$ & $<47.97$ & $<47.79$ & $<47.61$ & $<47.42$ & $<47.24$ & $<47.06$ & $<46.88$ & $<46.70$ & $<46.52$ \\
 HD 93083     & $50.62^{+0.62}_{-0.61}$ & $50.46^{+0.60}_{-0.60}$ & $50.29^{+0.59}_{-0.58}$ & $50.12^{+0.57}_{-0.57}$ & $49.96^{+0.55}_{-0.56}$ & $49.79^{+0.54}_{-0.54}$ & $49.63^{+0.52}_{-0.53}$ & $49.46^{+0.51}_{-0.51}$ & $49.29^{+0.50}_{-0.49}$ & $49.13^{+0.48}_{-0.48}$ & $48.96^{+0.47}_{-0.46}$ & $48.80^{+0.45}_{-0.45}$ & $48.63^{+0.44}_{-0.43}$ & $48.46^{+0.42}_{-0.42}$ & $48.30^{+0.40}_{-0.41}$ & $48.13^{+0.39}_{-0.39}$ & $47.97^{+0.37}_{-0.38}$ & $47.80^{+0.36}_{-0.36}$ \\
HD 95089      & $<50.79$ & $<50.61$ & $<50.43$ & $<50.24$ & $<50.06$ & $<49.88$ & $<49.70$ & $<49.52$ & $<49.34$ & $<49.16$ & $<48.98$ & $<48.80$ & $<48.62$ & $<48.43$ & $<48.25$ & $<48.07$ & $<47.89$ & $<47.71$ \\
 HD 99492     & $50.11^{+0.62}_{-0.61}$ & $49.95^{+0.60}_{-0.60}$ & $49.78^{+0.59}_{-0.59}$ & $49.61^{+0.57}_{-0.57}$ & $49.45^{+0.55}_{-0.56}$ & $49.28^{+0.54}_{-0.54}$ & $49.12^{+0.52}_{-0.53}$ & $48.95^{+0.51}_{-0.51}$ & $48.78^{+0.50}_{-0.49}$ & $48.62^{+0.48}_{-0.48}$ & $48.45^{+0.47}_{-0.46}$ & $48.29^{+0.45}_{-0.45}$ & $48.12^{+0.44}_{-0.43}$ & $47.95^{+0.42}_{-0.42}$ & $47.79^{+0.40}_{-0.41}$ & $47.62^{+0.39}_{-0.39}$ & $47.46^{+0.37}_{-0.38}$ & $47.29^{+0.36}_{-0.36}$ \\
HD 101930     & $49.13^{+0.62}_{-0.61}$ & $48.97^{+0.60}_{-0.60}$ & $48.80^{+0.59}_{-0.58}$ & $48.63^{+0.57}_{-0.57}$ & $48.47^{+0.55}_{-0.56}$ & $48.30^{+0.54}_{-0.54}$ & $48.14^{+0.52}_{-0.53}$ & $47.97^{+0.51}_{-0.51}$ & $47.80^{+0.50}_{-0.49}$ & $47.64^{+0.48}_{-0.48}$ & $47.47^{+0.47}_{-0.46}$ & $47.31^{+0.45}_{-0.45}$ & $47.14^{+0.44}_{-0.43}$ & $46.97^{+0.42}_{-0.42}$ & $46.81^{+0.40}_{-0.41}$ & $46.64^{+0.39}_{-0.39}$ & $46.48^{+0.37}_{-0.38}$ & $46.31^{+0.36}_{-0.36}$ \\
 HD 102195    & $51.92^{+0.62}_{-0.61}$ & $51.76^{+0.60}_{-0.60}$ & $51.59^{+0.59}_{-0.58}$ & $51.42^{+0.57}_{-0.57}$ & $51.26^{+0.55}_{-0.56}$ & $51.09^{+0.54}_{-0.54}$ & $50.93^{+0.52}_{-0.53}$ & $50.76^{+0.51}_{-0.51}$ & $50.59^{+0.50}_{-0.49}$ & $50.43^{+0.48}_{-0.48}$ & $50.26^{+0.47}_{-0.46}$ & $50.10^{+0.45}_{-0.45}$ & $49.93^{+0.44}_{-0.43}$ & $49.76^{+0.42}_{-0.42}$ & $49.60^{+0.40}_{-0.41}$ & $49.43^{+0.39}_{-0.39}$ & $49.27^{+0.37}_{-0.38}$ & $49.10^{+0.36}_{-0.36}$ \\
HD 108147     & $50.93^{+0.62}_{-0.61}$ & $50.77^{+0.60}_{-0.60}$ & $50.60^{+0.59}_{-0.59}$ & $50.43^{+0.57}_{-0.57}$ & $50.27^{+0.55}_{-0.56}$ & $50.10^{+0.54}_{-0.54}$ & $49.94^{+0.52}_{-0.53}$ & $49.77^{+0.51}_{-0.51}$ & $49.60^{+0.50}_{-0.49}$ & $49.44^{+0.48}_{-0.48}$ & $49.27^{+0.47}_{-0.46}$ & $49.11^{+0.45}_{-0.45}$ & $48.94^{+0.44}_{-0.43}$ & $48.77^{+0.42}_{-0.42}$ & $48.61^{+0.40}_{-0.41}$ & $48.44^{+0.39}_{-0.39}$ & $48.28^{+0.37}_{-0.38}$ & $48.11^{+0.36}_{-0.36}$ \\
 HD 111232    & $<50.05$ & $<49.87$ & $<49.69$ & $<49.50$ & $<49.32$ & $<49.14$ & $<48.96$ & $<48.78$ & $<48.60$ & $<48.42$ & $<48.24$ & $<48.06$ & $<47.88$ & $<47.69$ & $<47.51$ & $<47.33$ & $<47.15$ & $<46.97$ \\
 HD 114386    & $49.58^{+0.62}_{-0.61}$ & $49.42^{+0.60}_{-0.60}$ & $49.25^{+0.59}_{-0.58}$ & $49.08^{+0.57}_{-0.57}$ & $48.92^{+0.55}_{-0.56}$ & $48.75^{+0.54}_{-0.54}$ & $48.59^{+0.52}_{-0.53}$ & $48.42^{+0.51}_{-0.51}$ & $48.25^{+0.50}_{-0.49}$ & $48.09^{+0.48}_{-0.48}$ & $47.92^{+0.47}_{-0.46}$ & $47.76^{+0.45}_{-0.45}$ & $47.59^{+0.44}_{-0.43}$ & $47.42^{+0.42}_{-0.42}$ & $47.26^{+0.40}_{-0.41}$ & $47.09^{+0.39}_{-0.39}$ & $46.93^{+0.37}_{-0.38}$ & $46.76^{+0.36}_{-0.36}$ \\
HD 114762     & $<50.19$ & $<50.01$ & $<49.83$ & $<49.64$ & $<49.46$ & $<49.28$ & $<49.10$ & $<48.92$ & $<48.74$ & $<48.56$ & $<48.38$ & $<48.20$ & $<48.01$ & $<47.83$ & $<47.65$ & $<47.47$ & $<47.29$ & $<47.11$ \\
 HD 114783    & $<50.21$ & $<50.03$ & $<49.85$ & $<49.66$ & $<49.48$ & $<49.30$ & $<49.12$ & $<48.94$ & $<48.76$ & $<48.58$ & $<48.40$ & $<48.22$ & $<48.03$ & $<47.85$ & $<47.67$ & $<47.49$ & $<47.31$ & $<47.13$ \\
HD 130322     & $50.82^{+0.62}_{-0.61}$ & $50.66^{+0.60}_{-0.60}$ & $50.49^{+0.59}_{-0.59}$ & $50.32^{+0.57}_{-0.57}$ & $50.16^{+0.55}_{-0.56}$ & $49.99^{+0.54}_{-0.54}$ & $49.83^{+0.52}_{-0.53}$ & $49.66^{+0.51}_{-0.51}$ & $49.49^{+0.50}_{-0.49}$ & $49.33^{+0.48}_{-0.48}$ & $49.16^{+0.47}_{-0.46}$ & $49.00^{+0.45}_{-0.45}$ & $48.83^{+0.43}_{-0.43}$ & $48.66^{+0.42}_{-0.42}$ & $48.50^{+0.40}_{-0.41}$ & $48.33^{+0.39}_{-0.39}$ & $48.17^{+0.37}_{-0.38}$ & $48.00^{+0.36}_{-0.36}$ \\
 HD 154345    & $50.05^{+0.62}_{-0.61}$ & $49.89^{+0.60}_{-0.60}$ & $49.72^{+0.59}_{-0.59}$ & $49.55^{+0.57}_{-0.57}$ & $49.39^{+0.55}_{-0.56}$ & $49.22^{+0.54}_{-0.54}$ & $49.06^{+0.52}_{-0.53}$ & $48.89^{+0.51}_{-0.51}$ & $48.72^{+0.50}_{-0.49}$ & $48.56^{+0.48}_{-0.48}$ & $48.39^{+0.47}_{-0.46}$ & $48.23^{+0.45}_{-0.45}$ & $48.06^{+0.43}_{-0.44}$ & $47.89^{+0.42}_{-0.42}$ & $47.73^{+0.40}_{-0.41}$ & $47.56^{+0.39}_{-0.39}$ & $47.40^{+0.37}_{-0.38}$ & $47.23^{+0.36}_{-0.36}$ \\
 HD 164922    & $<49.44$ & $<49.26$ & $<49.08$ & $<48.89$ & $<48.71$ & $<48.53$ & $<48.35$ & $<48.17$ & $<47.99$ & $<47.81$ & $<47.63$ & $<47.45$ & $<47.26$ & $<47.08$ & $<46.90$ & $<46.72$ & $<46.54$ & $<46.36$ \\
HD 179949     & $51.65^{+0.62}_{-0.61}$ & $51.49^{+0.60}_{-0.60}$ & $51.32^{+0.59}_{-0.58}$ & $51.15^{+0.57}_{-0.57}$ & $50.99^{+0.55}_{-0.56}$ & $50.82^{+0.54}_{-0.54}$ & $50.66^{+0.52}_{-0.53}$ & $50.49^{+0.51}_{-0.51}$ & $50.32^{+0.50}_{-0.49}$ & $50.16^{+0.48}_{-0.48}$ & $49.99^{+0.47}_{-0.46}$ & $49.83^{+0.45}_{-0.45}$ & $49.66^{+0.44}_{-0.43}$ & $49.49^{+0.42}_{-0.42}$ & $49.33^{+0.40}_{-0.41}$ & $49.16^{+0.39}_{-0.39}$ & $49.00^{+0.37}_{-0.38}$ & $48.83^{+0.36}_{-0.36}$ \\
HD 187123     & $<50.98$ & $<50.80$ & $<50.62$ & $<50.43$ & $<50.25$ & $<50.07$ & $<49.89$ & $<49.71$ & $<49.53$ & $<49.35$ & $<49.17$ & $<48.99$ & $<48.81$ & $<48.62$ & $<48.44$ & $<48.26$ & $<48.08$ & $<47.90$ \\
HD 189733     & $50.10^{+0.69}_{-0.69}$ & $49.95^{+0.66}_{-0.67}$ & $49.80^{+0.64}_{-0.64}$ & $49.64^{+0.62}_{-0.61}$ & $49.49^{+0.59}_{-0.59}$ & $49.34^{+0.56}_{-0.57}$ & $49.18^{+0.54}_{-0.53}$ & $49.03^{+0.51}_{-0.51}$ & $48.88^{+0.48}_{-0.49}$ & $48.72^{+0.46}_{-0.45}$ & $48.57^{+0.43}_{-0.43}$ & $48.42^{+0.40}_{-0.41}$ & $48.27^{+0.38}_{-0.38}$ & $48.11^{+0.36}_{-0.35}$ & $47.96^{+0.33}_{-0.33}$ & $47.81^{+0.30}_{-0.31}$ & $47.65^{+0.28}_{-0.27}$ & $47.50^{+0.25}_{-0.25}$ \\
HD 190360     & $<50.05$ & $<49.87$ & $<49.69$ & $<49.50$ & $<49.32$ & $<49.14$ & $<48.96$ & $<48.78$ & $<48.60$ & $<48.42$ & $<48.24$ & $<48.06$ & $<47.88$ & $<47.69$ & $<47.51$ & $<47.33$ & $<47.15$ & $<46.97$ \\
HD 195019     & $<49.91$ & $<49.73$ & $<49.55$ & $<49.36$ & $<49.18$ & $<49.00$ & $<48.82$ & $<48.64$ & $<48.46$ & $<48.28$ & $<48.10$ & $<47.92$ & $<47.74$ & $<47.55$ & $<47.37$ & $<47.19$ & $<47.01$ & $<46.83$ \\
HD 209458     & $<50.10$ & $<49.92$ & $<49.74$ & $<49.55$ & $<49.37$ & $<49.19$ & $<49.01$ & $<48.83$ & $<48.65$ & $<48.47$ & $<48.29$ & $<48.11$ & $<47.92$ & $<47.74$ & $<47.56$ & $<47.38$ & $<47.20$ & $<47.02$ \\
HD 216435     & $51.24^{+0.62}_{-0.61}$ & $51.08^{+0.60}_{-0.60}$ & $50.91^{+0.59}_{-0.58}$ & $50.74^{+0.57}_{-0.57}$ & $50.58^{+0.55}_{-0.56}$ & $50.41^{+0.54}_{-0.54}$ & $50.25^{+0.52}_{-0.53}$ & $50.08^{+0.51}_{-0.51}$ & $49.91^{+0.50}_{-0.49}$ & $49.75^{+0.48}_{-0.48}$ & $49.58^{+0.47}_{-0.46}$ & $49.42^{+0.45}_{-0.45}$ & $49.25^{+0.44}_{-0.43}$ & $49.08^{+0.42}_{-0.42}$ & $48.92^{+0.40}_{-0.41}$ & $48.75^{+0.39}_{-0.39}$ & $48.59^{+0.37}_{-0.38}$ & $48.42^{+0.36}_{-0.36}$ \\
HD 216437     & $49.66^{+0.62}_{-0.61}$ & $49.50^{+0.60}_{-0.60}$ & $49.33^{+0.59}_{-0.59}$ & $49.16^{+0.57}_{-0.57}$ & $49.00^{+0.55}_{-0.56}$ & $48.83^{+0.54}_{-0.54}$ & $48.67^{+0.52}_{-0.53}$ & $48.50^{+0.51}_{-0.51}$ & $48.33^{+0.50}_{-0.49}$ & $48.17^{+0.48}_{-0.48}$ & $48.00^{+0.47}_{-0.46}$ & $47.84^{+0.45}_{-0.45}$ & $47.67^{+0.43}_{-0.43}$ & $47.50^{+0.42}_{-0.42}$ & $47.34^{+0.40}_{-0.41}$ & $47.17^{+0.39}_{-0.39}$ & $47.01^{+0.37}_{-0.38}$ & $46.84^{+0.36}_{-0.36}$ \\
HD 217107     & $<49.20$ & $<49.02$ & $<48.84$ & $<48.65$ & $<48.47$ & $<48.29$ & $<48.11$ & $<47.93$ & $<47.75$ & $<47.57$ & $<47.39$ & $<47.21$ & $<47.03$ & $<46.84$ & $<46.66$ & $<46.48$ & $<46.30$ & $<46.12$ \\
HD 218566     & $50.60^{+0.62}_{-0.61}$ & $50.44^{+0.60}_{-0.60}$ & $50.27^{+0.59}_{-0.58}$ & $50.10^{+0.57}_{-0.57}$ & $49.94^{+0.55}_{-0.56}$ & $49.77^{+0.54}_{-0.54}$ & $49.61^{+0.52}_{-0.53}$ & $49.44^{+0.51}_{-0.51}$ & $49.27^{+0.50}_{-0.49}$ & $49.11^{+0.48}_{-0.48}$ & $48.94^{+0.47}_{-0.46}$ & $48.78^{+0.45}_{-0.45}$ & $48.61^{+0.44}_{-0.43}$ & $48.44^{+0.42}_{-0.42}$ & $48.28^{+0.40}_{-0.41}$ & $48.11^{+0.39}_{-0.39}$ & $47.95^{+0.37}_{-0.38}$ & $47.78^{+0.36}_{-0.36}$ \\
HD 330075     & $49.60^{+0.62}_{-0.61}$ & $49.44^{+0.60}_{-0.60}$ & $49.27^{+0.59}_{-0.58}$ & $49.10^{+0.57}_{-0.57}$ & $48.94^{+0.55}_{-0.56}$ & $48.77^{+0.54}_{-0.54}$ & $48.61^{+0.52}_{-0.53}$ & $48.44^{+0.51}_{-0.51}$ & $48.27^{+0.50}_{-0.49}$ & $48.11^{+0.48}_{-0.48}$ & $47.94^{+0.47}_{-0.46}$ & $47.78^{+0.45}_{-0.45}$ & $47.61^{+0.44}_{-0.43}$ & $47.44^{+0.42}_{-0.42}$ & $47.28^{+0.40}_{-0.41}$ & $47.11^{+0.39}_{-0.39}$ & $46.95^{+0.37}_{-0.38}$ & $46.78^{+0.36}_{-0.36}$ \\
HR 8799       & $51.63^{+0.62}_{-0.61}$ & $51.47^{+0.60}_{-0.60}$ & $51.30^{+0.59}_{-0.58}$ & $51.13^{+0.57}_{-0.57}$ & $50.97^{+0.55}_{-0.56}$ & $50.80^{+0.54}_{-0.54}$ & $50.64^{+0.52}_{-0.53}$ & $50.47^{+0.51}_{-0.51}$ & $50.30^{+0.50}_{-0.49}$ & $50.14^{+0.48}_{-0.48}$ & $49.97^{+0.47}_{-0.46}$ & $49.81^{+0.45}_{-0.45}$ & $49.64^{+0.44}_{-0.43}$ & $49.47^{+0.42}_{-0.42}$ & $49.31^{+0.40}_{-0.41}$ & $49.14^{+0.39}_{-0.39}$ & $48.98^{+0.37}_{-0.38}$ & $48.81^{+0.36}_{-0.36}$ \\
$\mu$ Ara    & $<49.68$ & $<49.50$ & $<49.32$ & $<49.13$ & $<48.95$ & $<48.77$ & $<48.59$ & $<48.41$ & $<48.23$ & $<48.05$ & $<47.87$ & $<47.69$ & $<47.51$ & $<47.32$ & $<47.14$ & $<46.96$ & $<46.78$ & $<46.60$ \\
NGC 2423 3    & $<57.27$ & $<57.09$ & $<56.91$ & $<56.72$ & $<56.54$ & $<56.36$ & $<56.18$ & $<56.00$ & $<55.82$ & $<55.64$ & $<55.46$ & $<55.28$ & $<55.10$ & $<54.91$ & $<54.73$ & $<54.55$ & $<54.37$ & $<54.19$ \\
Pollux        & $50.28^{+0.62}_{-0.61}$ & $50.12^{+0.60}_{-0.60}$ & $49.95^{+0.59}_{-0.58}$ & $49.78^{+0.57}_{-0.57}$ & $49.62^{+0.55}_{-0.56}$ & $49.45^{+0.54}_{-0.54}$ & $49.29^{+0.52}_{-0.53}$ & $49.12^{+0.51}_{-0.51}$ & $48.95^{+0.50}_{-0.49}$ & $48.79^{+0.48}_{-0.48}$ & $48.62^{+0.47}_{-0.46}$ & $48.46^{+0.45}_{-0.45}$ & $48.29^{+0.44}_{-0.43}$ & $48.12^{+0.42}_{-0.42}$ & $47.96^{+0.40}_{-0.41}$ & $47.79^{+0.39}_{-0.39}$ & $47.63^{+0.37}_{-0.38}$ & $47.46^{+0.36}_{-0.36}$ \\
$\tau$ Boo\tablefootmark{b}   & $51.82^{+0.62}_{-0.61}$ & $51.66^{+0.60}_{-0.60}$ & $51.49^{+0.59}_{-0.59}$ & $51.32^{+0.57}_{-0.57}$ & $51.16^{+0.55}_{-0.56}$ & $50.99^{+0.54}_{-0.54}$ & $50.83^{+0.52}_{-0.53}$ & $50.66^{+0.51}_{-0.51}$ & $50.49^{+0.50}_{-0.49}$ & $50.33^{+0.48}_{-0.48}$ & $50.16^{+0.47}_{-0.46}$ & $50.00^{+0.45}_{-0.45}$ & $49.83^{+0.43}_{-0.43}$ & $49.66^{+0.42}_{-0.42}$ & $49.50^{+0.40}_{-0.41}$ & $49.33^{+0.39}_{-0.39}$ & $49.17^{+0.37}_{-0.38}$ & $49.00^{+0.36}_{-0.36}$ \\
$\upsilon$ And& $50.97^{+0.62}_{-0.61}$ & $50.81^{+0.60}_{-0.60}$ & $50.64^{+0.59}_{-0.58}$ & $50.47^{+0.57}_{-0.57}$ & $50.31^{+0.55}_{-0.56}$ & $50.14^{+0.54}_{-0.54}$ & $49.98^{+0.52}_{-0.53}$ & $49.81^{+0.51}_{-0.51}$ & $49.64^{+0.50}_{-0.49}$ & $49.48^{+0.48}_{-0.48}$ & $49.31^{+0.47}_{-0.46}$ & $49.15^{+0.45}_{-0.45}$ & $48.98^{+0.44}_{-0.43}$ & $48.81^{+0.42}_{-0.42}$ & $48.65^{+0.40}_{-0.41}$ & $48.48^{+0.39}_{-0.39}$ & $48.32^{+0.37}_{-0.38}$ & $48.15^{+0.36}_{-0.36}$ \\
%---------------
\hline
\end{tabular}
\end{scriptsize}
%\end{center}
\tablefoot{
\tablefoottext{a}{Emission measure (EM=log $\int N_{\rm e} N_{\rm H}
  {\rm d}V$), where $N_{\rm e}$ 
and $N_{\rm H}$ are electron and hydrogen densities, in
cm$^{-3}$. }
\tablefoottext{b}{EMD at coronal temperatures from \citet{mag11}. }
}.
\end{table}
\end{landscape}
%--------------------------------------------- end table

%% file: fluxesacen.tex
%\onltab{2}{
%------------------------------------------------- begin table
% XMM/RGS alpha Cen lines table
\begin{table}
\caption{XMM/RGS line fluxes of $\alpha$~Cen B\tablefootmark{a}}\label{tab:acenfluxes}
\tabcolsep 3.pt
\begin{scriptsize}
\begin{tabular}{lrcrrrl}
\hline \hline
 Ion & {$\lambda$$_{\mathrm {model}}$} &  
 log $T_{\mathrm {max}}$ & $F_{\mathrm {obs}}$ & S/N & ratio & Blends \\ 
%     & (\AA) & (K) & erg\,cm$^{-2}$\,s${-1}$ & & & \\ 
\hline
%------------------------------------------------------
\ion{Ne}{ix} & 13.4473 & 6.6 & 9.59e-15 &   5.6 & -0.10 & \ion{Fe}{xix} 13.4970, 13.5180, \\
               &         &     &          &       &       & \ion{Ne}{ix} 13.5531  \\
\ion{Ne}{ix} & 13.6990 & 6.6 & 1.30e-14 &   5.3 &  0.08 & \ion{Ni}{xix} 13.7790, \ion{Fe}{xvii} 13.8250 \\
\ion{Fe}{xvii} & 15.0140 & 6.7 & 6.62e-14 &  10.7 & -0.03 &  \\
\ion{Fe}{xvii} & 15.2610 & 6.7 & 4.29e-14 &   8.1 &  0.20 & \ion{O }{viii} 15.1760, \\
               &         &     &          &       &       & \ion{Fe}{xvii} 15.2509, 15.2615 \\
\ion{O }{viii} & 16.0055 & 6.5 & 1.23e-14 &   4.0 & -0.02 & \ion{Fe}{xvii} 15.9956, \ion{Fe}{xviii} 16.0040, \\ 
               &         &     &          &       &       & \ion{O }{viii} 16.0067 \\
\ion{Fe}{xvii} & 16.7800 & 6.7 & 1.38e-14 &   3.9 & -0.35 &  \\
\ion{Fe}{xvii} & 17.0510 & 6.7 & 6.35e-14 &   9.8 & -0.04 & \ion{Fe}{xvii} 17.0960 \\
\ion{O }{vii} & 18.6270 & 6.3 & 1.16e-14 &   3.1 & -0.10 &  \\
\ion{O }{viii} & 18.9671 & 6.5 & 7.65e-14 &  10.6 &  0.00 & \ion{O }{viii} 18.9725 \\
\ion{O }{vii} & 21.6015 & 6.3 & 1.29e-13 &  20.2 &  0.02 &  \\
\ion{O }{vii} & 22.0977 & 6.3 & 1.22e-13 &  12.5 &  0.19 &  \\
\ion{N }{vii} & 24.7792 & 6.3 & 3.68e-14 &  11.1 & -0.04 & \ion{N }{vii} 24.7846 \\
\ion{Ca}{xi} & 25.3520 & 6.3 & 7.10e-15 &   3.0 &  0.28 & \ion{N }{vii} 25.4030 \\
\ion{C }{vi} & 28.4652 & 6.2 & 1.32e-14 &   3.3 & -0.27 & \ion{Ar}{xv} 28.3860, \ion{C }{vi} 28.4663 \\
\ion{N }{vi} & 28.7870 & 6.2 & 2.54e-14 &   5.0 & -0.08 &  \\
\ion{N }{vi} & 29.5347 & 6.1 & 2.69e-14 &   5.5 &  0.13 &  \\
\ion{Ca}{xi} & 30.4710 & 6.3 & 2.51e-14 &   4.7 & -0.14 & \ion{S }{xiv} 30.4270, 30.4690 \\
\ion{C }{vi} & 33.7342 & 6.1 & 1.83e-13 &  12.4 &  0.01 & \ion{C }{vi} 33.7396 \\
\ion{C }{v} & 34.9728 & 6.0 & 1.49e-14 &   3.2 &  0.03 & \ion{Ar}{ix} 35.0240 \\
\ion{Ca}{xi} & 35.2750 & 6.3 & 1.87e-14 &   3.6 &  0.16 & \ion{S }{xii} 35.2750 \\
\ion{S }{xiii} & 35.6670 & 6.4 & 3.93e-14 &   5.6 & -0.01 & \ion{Ca}{xi} 35.6340, 35.7370 \\
%-----------------
\hline
\end{tabular}
\tablefoot{\tablefoottext{a}{Line fluxes in erg cm$^{-2}$ s$^{-1}$. $\lambda$$_{\mathrm {model}}$ (\AA) is the APED model wavelength 
  corresponding to the measured line. 
  $\log T_{\rm max}$ indicates the maximum
  temperature (K) of formation of the line (unweighted by the
  EMD). ``Ratio'' is the $\log (F_{\mathrm {obs}} / F_{\mathrm {pred}})$ 
  of the line. 
  Blends amounting to more than 5\% of the total flux for each line are
  indicated.}}
\end{scriptsize}
\end{table}
%-----------------